\documentclass[prb,aps,amsfonts,amsmath,amssymb,nofootinbib,twocolumn,superscriptaddress]{revtex4}
\usepackage{graphicx}
\usepackage{bm}
\usepackage{hyperref}
\usepackage{IEEEtrantools}
\usepackage{natbib}
\usepackage{float}
\usepackage{siunitx}
\usepackage{enumitem}
\usepackage{xcolor}
\usepackage{graphicx,mathtools}

\restylefloat{table}

\begin{document}

\title{Observation of a marginal Fermi glass using THz 2D coherent spectroscopy}

\author{Fahad Mahmood}
\affiliation{Department of Physics and Astronomy, The Johns Hopkins University, Baltimore, Maryland 21218, USA}
\affiliation{Department of Physics, University of Illinois at Urbana-Champaign, Urbana, 61801 IL, USA}
\affiliation{F. Seitz Materials Research Laboratory, University of Illinois at Urbana-Champaign, Urbana, 61801 IL, USA}

\author{Dipanjan Chaudhuri}
\affiliation{Department of Physics and Astronomy, The Johns Hopkins University, Baltimore, Maryland 21218, USA}

\author{Sarang Gopalakrishnan}
\affiliation{Department of Engineering Science and Physics, CUNY College of Staten Island, Staten Island, NY 10314, USA}
\affiliation{Initiative for Theoretical Sciences, The Graduate Center, CUNY, New York, NY 10016, USA}

\author{Rahul Nandkishore}
\affiliation{Department of Physics and Center for Theory of Quantum Matter, University of Colorado Boulder, Boulder, Colorado 80309, USA}

\author{N. P. Armitage}\email{npa@jhu.edu}
\affiliation{Department of Physics and Astronomy, The Johns Hopkins University, Baltimore, Maryland 21218, USA}

\maketitle

\textbf{A longstanding open problem in condensed matter physics is whether or not a strongly disordered interacting insulator can be mapped to a system of effectively non-interacting localized excitations. We investigate this issue on the insulating side of the 3D metal-insulator transition (MIT) in phosphorus doped silicon using the new technique of terahertz two dimensional coherent spectroscopy. Despite the intrinsically disordered nature of these materials, we observe coherent excitations and strong photon echoes that provide us with a powerful method for the study of their decay processes.  We extract the first measurements of energy relaxation ($T_1$) and decoherence ($T_2$)  times close to the MIT in this classic system.  We observe that (i) both relaxation rates are linear in excitation frequency with a slope close to unity, (ii) the energy relaxation timescale $T_1$ counterintuitively \textit{increases} with increasing temperature and (iii) the coherence relaxation timescale $T_2$ has little temperature dependence between $\SI{5}{K}$ and $\SI{25}{K}$, but counterintuitively \textit{increases} as the material is doped towards the MIT. We argue that these features imply that (a) the system behaves as a well isolated electronic system on the timescales of interest, and (b) relaxation is controlled by electron-electron interactions. We discuss the potential relaxation channels that may explain the behavior. Our observations constitute a qualitatively new phenomenology, driven by the interplay of strong disorder and strong electron-electron interactions, which we dub the marginal Fermi glass.}

\bigskip

Understanding systems with strong disorder and strong interactions is a central open issue in condensed matter physics.  It is a remarkable fact that many {\it metals} can be understood in terms of weakly interacting fermionic quasiparticles near the Fermi energy ($E_F$) despite the fact that the bare Coulomb interaction is not particularly small or short-ranged. This has been canonized in terms of the Landau Fermi liquid theory \cite{nozierespines}, where the effects of interactions renormalize quasiparticle parameters like the effective mass, but do not change the underlying effective structure of the theory from that of free-electrons.  Scattering rates of the quasiparticles in a Landau Fermi liquid go like $(E-E_F)^2$; thus quasiparticles are arbitrarily well-defined near $E_F$.  These effects arise as a consequence of both the Pauli exclusion principle, which reduces the phase space for scattering, and screening that renders the bare Coulomb interaction effectively short-range.

The tendency of strong disorder is to localize particles. Anderson showed that in the absence of interactions sufficiently strong disorder could localize wavefunctions with a sharp boundary in energy between localized and extended states~\cite{pwa1958}. This is a generic wave phenomenon that applies equally to acoustic, electromagnetic, or neutral matter waves~\cite{abrahams201050}. Although such ``Anderson localization'' is frequently invoked in the study of disordered electronic insulators, it is unclear to what extent this phenomenon actually applies to real materials.

\begin{figure*}
	\includegraphics[width=1.9\columnwidth]{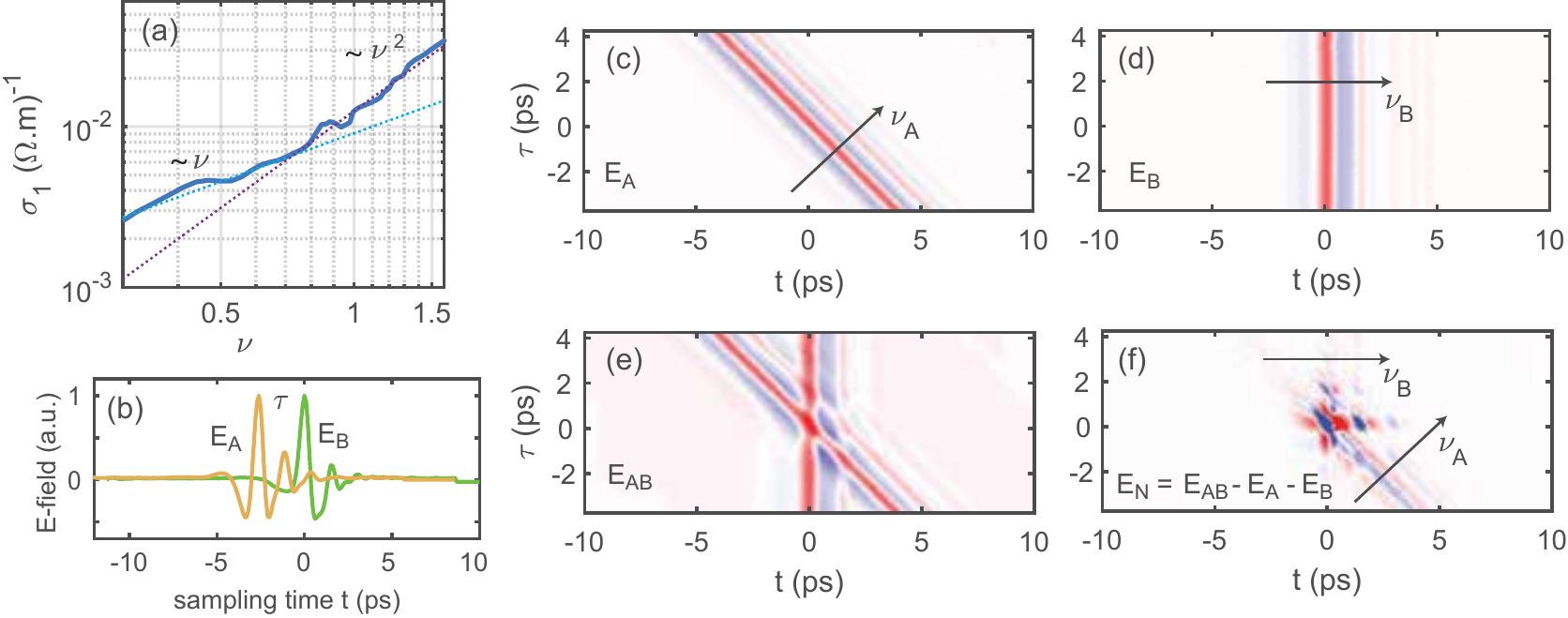}
	\caption{\textbf{Linear and nonlinear optical response of phosphorus doped silicon}  \textbf{(a)} Linear response optical conductivity for the 39$\%$ sample at $\SI{5}{K}$.   Regime of linear and quadratic power-law behavior can be distinguished (dashed lines).  \textbf{(b)} Time traces of two collinear THz pulses that are separated by a time $\tau$.   In the experiment, the sum of these electric fields $E_{AB}$ is measured.   \textbf{(c)-(f)} Time traces of  $E_A$, $E_B$, $E_{AB}$, and $E_{NL} = E_{AB}$ - $E_{A}$ - $E_{B}$ respectively as a function of $t$ and $\tau$.}\label{LinearTimeTraces}
\end{figure*}

In this regard, in 1970 Anderson proposed the notion  -- in analogy with the Fermi liquid -- of a ``Fermi glass" as a localized disordered state of matter adiabatically connected to the noninteracting Anderson insulator, whose universal properties arose through Pauli exclusion alone \cite{Anderson70a}.  Anderson conjectured that via the protection afforded by the Fermi energy, such a state of matter would also have well-defined single-particle-like excitations at low energy.  It was later understood that the localized nature of such systems and lack of metallic screening made these considerations more subtle \cite{Efros75a, fleishman1980interactions, freedman1977theory}.  Recently, it was realized that insulators might feature an even stronger notion of adiabatic continuity than metals.  Refs.~\cite{gmp, basko2006metal} argued that a disordered system with {\it short-range interactions} could be ``many-body-localized" and thus have infinitely sharp excitations even at nonzero temperatures and far from $E_F$.  This has been a large area of current inquiry---see \cite{mblarcmp, mblrmp} for reviews. 
However the effects of {\it long-range Coulomb interactions} are still not fully understood. It has been shown \cite{burin1998, burin, yao2014many, gutman} that long-range interactions invalidate perturbative arguments for localization.  While non-perturbative methods have been applied in certain settings \cite{lrmbl}, it remains unknown whether a ``Fermi glass'' exists, i.e., whether a frequency or temperature window exists where the excitations of an interacting insulator are renormalized, weakly interacting, electron-like quasiparticles. 

In this work we use the new technique of terahertz 2D coherent spectroscopy (THz 2DCS)\cite{woerner2013ultrafast,lu2017,wan2019resolving} to shed light on this fundamental problem.  We investigate the canonical disordered material phosphorus doped silicon (Si:P) on the insulating side of the metal-insulator transition \cite{rosenbaum1983metal,paalanen1983critical}. Among other aspects, THz 2DCS allows us to measure both $T_1$ and $T_2$ times of inhomogeneously broadened spectra in the THz range.  At low temperature we find a temperature independent regime governed by electron-electron interactions.  We find that relaxation rates of the optical excitations are linear in frequency with a proportionality constant of order one. This establishes that in our frequency range, the low energy excitations are not well defined.  This is consistent with a picture in which localized electronic systems are {\it not} adiabatically connected to the Anderson insulator.  We call this state of matter the {\it marginal Fermi glass}.  This electronic relaxation is consistent with the existence of an electronic continuum that arises through long-range Coulomb interactions which could destabilize the localized state at non-zero temperatures.

In Fig.~\ref{LinearTimeTraces}a, we show the real part of the linear response THz range conductivity of a representative 39$\%$ sample.   In previous work \cite{Helgren04a,Helgren02a} it was shown that the optical response of Si:P near the MIT was in accord with the theory of Mott-Efros-Shklovskii \cite{Lee01a,shklovskii1981phononless}. Here one models the excited states of the system as an ensemble of resonant pairs that can be mapped to a random ensemble of two-level systems (i.e., the ``pair approximation") that gives a conductivity $ \sigma_{1}(\omega) = \alpha e^{2}N_{0}^{2}\xi^{5}\omega[ln(2I_{0}/\hbar\omega)]^{4}[\hbar\omega +   U(r_{w})]$ e.g., an almost linear conductivity is found at low frequencies and a quadratic one at higher frequencies.  These power laws come from phase space considerations (see Supplementary Information (SI)). They crossover at an energy scale $U(r_{\omega}) =  e^{2}/\varepsilon_{1} r_{\omega}$ which is the attraction between an electron and hole in a dipolar excitation at a distance $r_\omega = \xi[ln(2I_{0}/\hbar\omega)]$.  Here $\alpha$ is a constant close to one, and $I_{0}$ is the pre-factor of the overlap integral (commonly taken to be the Bohr energy of the dopant).  One aspect not considered in the usual treatment is that each of the excitations that contributes to $\sigma(\omega)$ has a finite lifetime. The functional form of $\sigma(\omega)$ is insensitive to moderate level broadening, so it is uninformative about excitation lifetimes.  Quantifying homogeneous broadening due to quasiparticle decay in the face of overtly inhomogeneously broadened spectra is the principal difficulty in characterizing interactions in these systems.

\begin{figure*}
	\includegraphics[width=17cm]{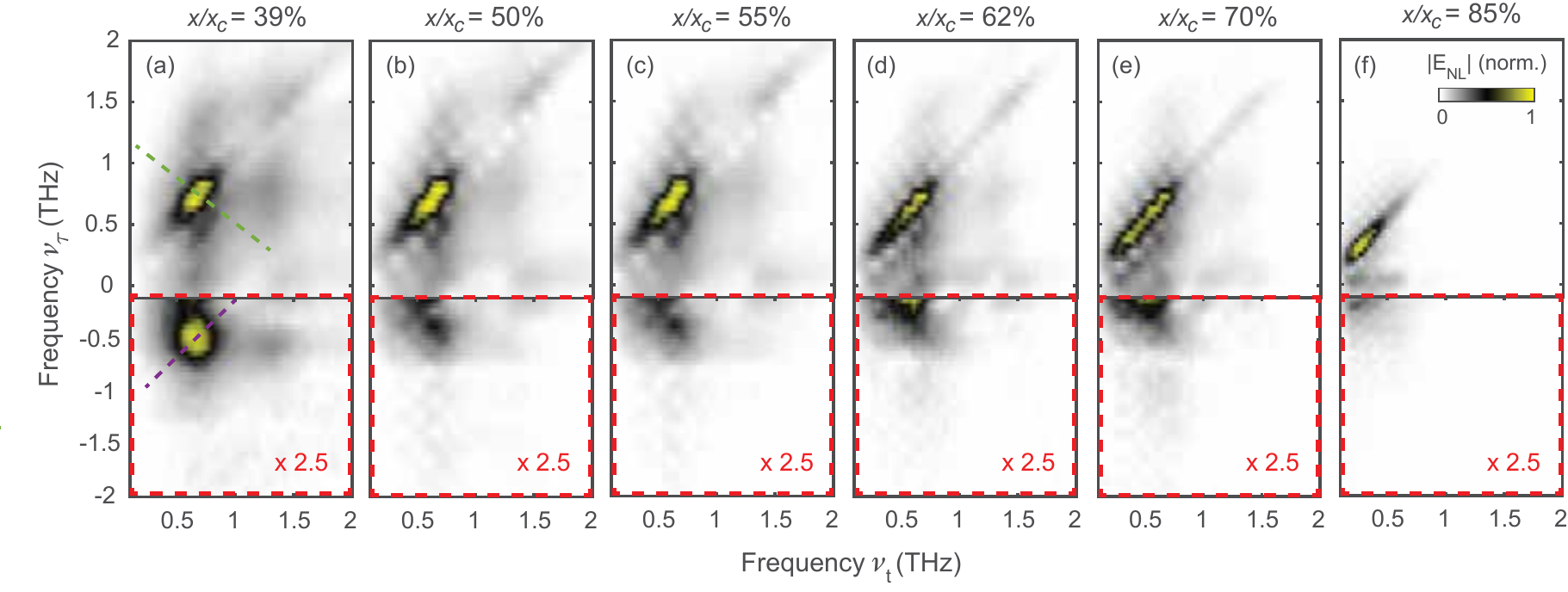}
	\caption{\label{2D spectrum} \textbf{2D THz spectra at different P concentrations.} 2D THz spectra $|E_{NL}(\nu_\tau,\nu_t)|$ at $T=\SI{5}{K}$ for a series of Si:P samples with different dopings.  The doping for each sample is expressed as a percent of the critical doping ($x_c$) at which the MIT occurs. The spectra are obtained by taking the absolute value of the 2D Fourier transform of the time domain signal $E_{NL}(t,\tau)$. The spectrum at each doping is normalized to its maximum and plotted according to the color map shown in (f). The intensity in the fourth quadrant (red dashed area) from which we get $\Gamma_2$ is magnified by x2.5.}
\end{figure*}

2D coherent spectroscopy is a nonlinear 4-wave mixing technique that can, among other aspects, directly reveal couplings between excitations and separate homogeneous from inhomogeneous broadening \cite{Mukamel1995,hamm2011concepts,aue1976two,cundiff2013optical}.  It has been incredibly powerful in its radio and infrared frequency incarnations for the study of chemical systems.  It has been extended recently to the THz range to study graphene and quantum wells~\cite{kuehn2011two,woerner2013ultrafast}, molecular rotations~\cite{lu2016nonlinear}, and spin waves in conventional magnets ~\cite{lu2017}. It has also been proposed to give unique information on fractionalized spin phases \cite{wan2019resolving,choi2020theory}.

As discussed in the method section, two THz pulses (A and B) are incident on a sample.  The transmitted electric field is recorded as a function of the separation between them ($\tau$) and the time from pulse B  ($t$).   The nonlinear signal is defined $E_{NL}(\tau,t) = E_{AB}(\tau,t) - E_{A}(\tau,t) - E_{B}(t)$.  Here $E_{AB}$ is the transmitted signal when both THz pulses are present while $E_{A}$ and $E_{B}$ are the transmitted signals with each pulse A and B present individually.  Fig.~1c-f shows $E_A$, $E_B$, $E_{AB}$, and $E_{NL}$ as a function of $t$ and $\tau$ for the 39\% sample.  Fig.~2 shows the resulting 2D THz spectra $E_{NL}(\nu_\tau,\nu_t)$ from the Fourier transforms with respect to $\tau$ and $t$ for each doping studied at a temperature of \SI{5}{K}. We note that typical P-P spacings in these samples are of order \SI{8}{nm} \cite{Helgren04a}, and hence the associated Coulomb energy is of order \SI{15}{meV}, which corresponds to a timescale of $\approx \SI{0.3}{ps}$. Thus, the experimental timescales of a few ps of these measurements are easily long enough for the effect of interactions to be important.

\begin{figure*}
	\includegraphics[width=17cm]{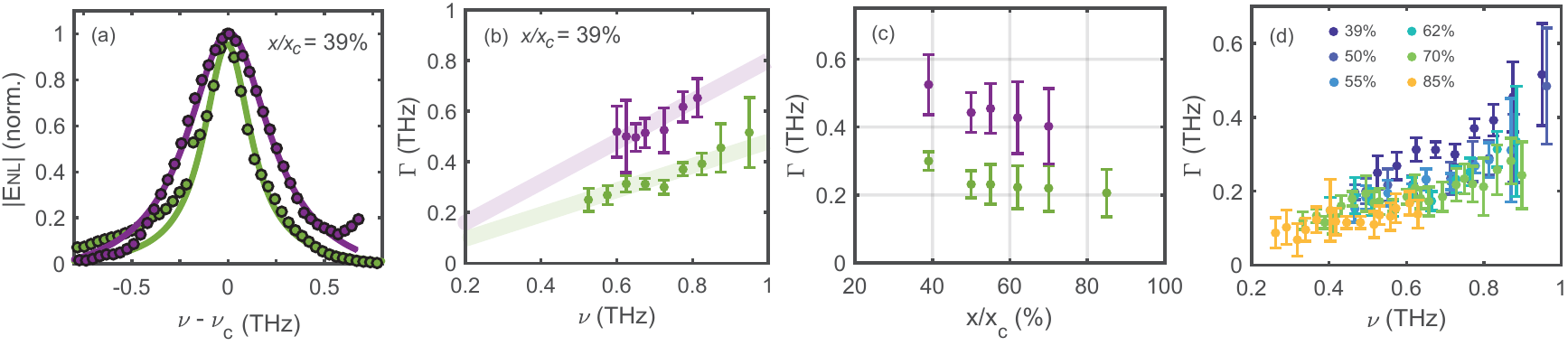}
	\caption{\label{frequencydependence} \textbf{Frequency and doping dependence of the pump-probe and rephasing relaxation rates.}  \textbf{(a)} $|E_{NL}|$ as a function of excitation frequency along the anti-diagonal cuts indicated by the dashed green and purple lines in Fig.~2a for the 39\% sample at \SI{5}{K}. Green dots are cut through the 1st quadrant PP signal while purple dots represent a cut through the 4th quadrant R signal. The solid lines show a Lorentzian best fit to the data. \textbf{(b)} As discussed in the SI, the widths of these two features can be interpreted as $\Gamma_1 = 1/T_1$ (green dots) and $\Gamma_2 = 1/T_2$ (purple dots) within a model of two levels systems. Shown as a function of excitation frequency for the 39\% sample. Light solid lines pass through the origin and are a linear best fit guide to the eye. \textbf{(c)} Relaxation rates $\Gamma_1$ (green dots) and $\Gamma_2$ (purple dots) at $\nu_t = $ \SI{0.67}{THz} as a function of doping. Error bars in (b) and (c) represent the 95\% confidence interval in the Lorentzian best fit. \textbf{(d)}  $\Gamma_1$ as a function of frequency at different dopings at \SI{5}{K}.}
\end{figure*}

In an inversion symmetric system like Si:P the leading nonlinear response is $\chi^{(3)}$ electric dipole reradiation.   Therefore with two pulses, there are contributions to $E_{NL}$ in which pulse A interacts twice and pulse B once with the sample and other contributions where pulse A interacts once and pulse B twice.  Moreover, the $\chi^{(3)}$ response can be separated into non-rephasing (NR) and rephasing (R) contributions \cite{woerner2013ultrafast,lu2017}.  The R signal arises due to a reverse phase accumulation during time $t$ as compared to $\tau$ and thus occurs at negative frequencies of either $\nu_t$ or $\nu\tau$ when compared with the NR signal.  The different non-linear signals in each quadrant in the 2D frequency plan can be understood in terms of `frequency vectors' as outlined in \cite{woerner2013ultrafast,wan2019resolving} and the SI.  For the case where pulse A precedes B and pulse B has two sample interactions (AB scheme), the $E_{NL}$ signal in the 4th quadrant is the `photon echo' R contribution.  Within a picture where excitations are resonant pairs, its anti-diagonal widths are a measure of the decoherence rates ($\Gamma_2 = 1/T_2$) \cite{Mukamel1995,hamm2011concepts,TomakoffNotes}.  The strong signal along the diagonal in the 1st quadrant is a pump-probe (PP) contribution from pulse B interacting twice from the sample and arriving before A (BA scheme).  It is sensitive to decay of the excited state populations and its anti-diagonal width is, within the pair approximation, a measure of the energy relaxation rate ($\Gamma_1 = 1/T_1$) at the excitation frequency of the projection onto either axis.  Please see the SI for a detailed description of the full 2DCS response of a generic two-level system subject to finite longitudinal and transverse relaxation rates $1/T_1$ and $1/T_2$.

\begin{figure*}
	\includegraphics[width=12cm]{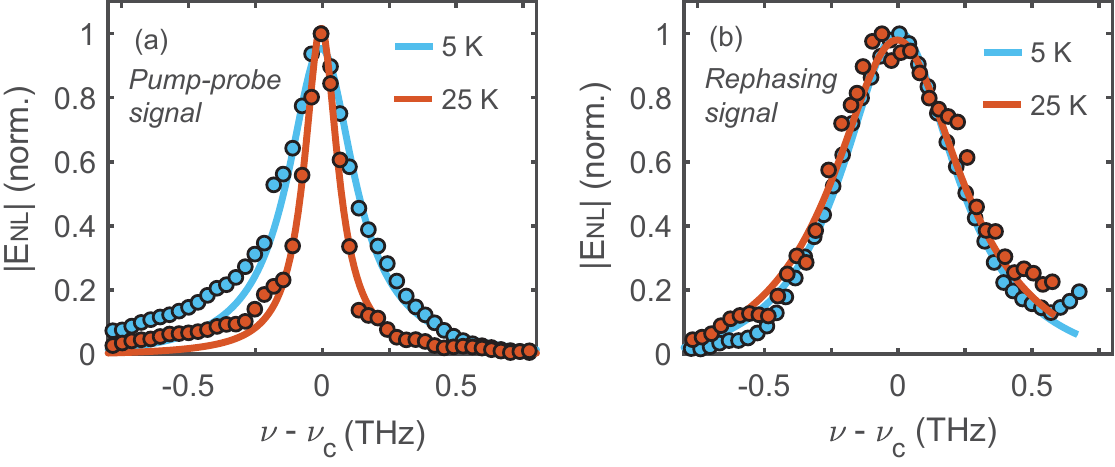}
	\caption{\label{tempdependence} \textbf{Temperature dependence of the  pump-probe and rephasing anti-diagonal spectra.}   \textbf{(a)} $|E_{NL}|$ as a function of frequency along the anti-diagonal cut of the PP signal in the 1st quadrant at $T= \SI{5}{K}$ and $T=\SI{25}{K}$ for the 39\% sample. \textbf{(b)} $|E_{NL}|$ as a function of frequency along the anti-diagonal cut of the R signal in the 4th quadrant at $T= \SI{5}{K}$ and $T=\SI{25}{K}$ for the 39\% sample.}
\end{figure*}

As can be seen in Fig. 2, the signals shift towards lower frequencies on approaching the MIT. The echo signal is most apparent for the least doped (39\%) sample (Fig. 2a). Similarly, the PP streak along the 1st quadrant diagonal  narrows (decreasing $\Gamma_{1}$) with increasing doping.  To quantify the relaxation rates $\Gamma_{1,2}$, we take cuts along the anti-diagonal in both quadrants. In so doing, we can get the relaxation rates as a function of energy.  Fig. \ref{frequencydependence}a shows representative cuts of the 39$\%$ sample taken along the green and purple dashed lines in Fig. 2a. The cuts in each case can be well fit to a single Lorentzian to extract the FWHM as a measure of the relaxation rates.  We plot the frequency dependencies of the relaxation rates in Fig.  \ref{frequencydependence}b and the doping dependencies in Fig.  \ref{frequencydependence}c.  The relative widths of  $\Gamma_{1,2}$ easily satisfy the fundamental relation for ``magnetic" resonance with $2/T_2 \geq 1/ T_1$.   The frequency dependence of $\Gamma_{1}$ is shown in Fig.  \ref{frequencydependence}d at different doping levels. Note that the x-axis in Fig.~3b and Fig.~3d is the frequency $\nu_t$ at which the anti-diagonal cuts peak in Fig.~2. Due to low signal it was challenging to extract $\Gamma_{2}$ over the full doping range for all samples.

One can see in Fig. \ref{frequencydependence}b and d that the relaxation rates are roughly linear in excitation frequency in the sub-THz regime and are consistent with an extrapolation to zero in the limit of zero frequency.  Qualitatively,  this frequency dependence is reminiscent of the behavior of the relaxation rate as a function of energy for a metal in the sense that they go to zero as $\omega \rightarrow  0$ (e.g., as $E \rightarrow E_F$) as the phase space for electronic relaxation collapses.   However, the quantitative dependence is different as the relaxation rates are linear in frequency with a slope close to unity.  We also find that the doping dependence (Fig. \ref{frequencydependence}c) is such that the relaxation rates {\it decrease} as we approach the transition to the metallic phase.  Fig. \ref{tempdependence} shows the temperature dependence of the relaxation rates. Over a range of temperatures from $5K$ to $25K$, $1/T_2$ does not change at all to within experimental uncertainty, while the $1/T_1$ relaxation rate actually {\it decreases} with increasing temperature.

What can be inferred from these results?  The  temperature dependence in Fig. \ref{tempdependence} and the frequency dependence in Fig. \ref{frequencydependence} rule out phonons as a dominant relaxation channel. Relaxation from phonons is known (see e.g. \cite{spectraldiffusion}), to lead to relaxation rates that are increasing functions of temperature and with frequency dependence  (see SI) that goes as $\omega^3$ at low $\omega$.  Thus, the electronic system can be considered well isolated on the timescales of interest with coupling to phonons unimportant for this relaxation. 
We further note that the temperature dependence of the energy relaxation rate $1/T_1$ (Fig.\ref{tempdependence}a) has the opposite sign from what one might naively expect - the relaxation rate {\it decreases} as we increase temperature.  This also rules out explanations based on spectral diffusion (see SI), but can be naturally explained if we postulate that $T_1$ comes from the interaction mediated coherent tunneling of electron-hole excitations (see SI).  Raising the temperature increases screening and suppresses coherent tunneling \cite{fisherzwerger}.

The doping dependence of the relaxation rate (Fig. \ref{frequencydependence}c) provides further insight. This behavior is counterintuitive: relaxation slows as we approach the metallic phase.  We argue that in fact, it is evidence that electron-electron interactions dominate relaxation.  This follows from essentially dimensional considerations. Microscopically, the system consists of randomly placed P atoms; electrons hop between these atoms, and repel each other via the Coulomb interaction. The system is doped toward the MIT by increasing the density of P atoms. Since the hopping is exponentially suppressed in the P-P distance and the Coulomb interactions are only algebraically suppressed, increasing the density \emph{decreases} the ratio of interactions to hopping, and thus makes the system effectively more weakly interacting, causing the quasiparticle lifetimes to increase.

The linear in frequency relaxation rate with a slope close to unity is a dependence reminiscent of strongly correlated metals that exhibit the marginal Fermi liquid phenomenology \cite{varma2002singular,varma1989phenomenology} and here demonstrates something similar; particle-hole excitations are only marginally well-defined in the relevant frequency range.   That the relaxation rates appear to extrapolate to zero in the zero frequency limit, despite the fact that the system is at finite temperature, likely reflects the fact that we are probing energy scales larger than the thermal scale; the smallest frequency probed ($\approx \SI{5}{THz}$) corresponds to a temperature of $\approx \SI{24}{K}$ and the data in Fig.\ref{frequencydependence}(a) is taken at $\SI{5}{K}$. It is likely that the relaxation rates saturate to a non-zero value at a frequency lower than we can probe due to the finite temperature of the experiment.

We now sketch a mechanism that can give the linear in $\omega$ dependence (see SI for details). The low-frequency excitations above the localized state are resonant particle-hole excitations, in which a particle is moved between two nearby localized orbitals. These excitations are local electric dipoles, and thus naturally interact via $1/R^3$ dipole-dipole interactions~\cite{burin1994low, gutman}. As discussed by Levitov~\cite{levitov1989absence}, dipolar interactions are known to cause delocalization in three dimensions. Coulomb interactions parametrically enhance the density of dipoles at low frequencies~\cite{shklovskii1981phononless}, through a blockade effect; even if two nearby sites have on-site energies below $E_F$, occupying one of them may push the other site above $E_F$. These local anticorrelations among occupation numbers give a phase space of particle hole excitations that is $\omega$-independent at low frequencies (cf. Fermi liquids, where this phase space scales as $\omega$).  A dipolar excitation can coherently hop on this network, at a rate one can calculate (see SI) to be $\sim \omega$.  This mechanism also has temperature and doping dependence consistent with the earlier dimensional analysis. It is important to point out that our experiment is not in the regime far from the MIT where the Shklovski-Efros-Levitov (S.E.L.) calculation is well controlled (see SI). Nevertheless it remains possible that the experimental results are quantitatively explicable via some nontrivial extension of its central ideas to systems near the MIT.  Regardless of the precise mechanism, the relaxation comes from the interplay of strong electron electron interactions with strong disorder, in a regime where controlled analytic calculation does not appear feasible.  It should be noted that in the absence of the Efros-Shklovskii ground state reconstruction \cite{shklovskii1981phononless}, the Levitov argument \cite{levitov1989absence} would predict relaxation rates that scale as $\sim \omega^2$, implying sharply defined low energy excitations (see SI).   One needs both it and the long-range dipolar interaction to get the linear in $\omega$ relaxation.

Finally, we note that while discussions of Fermi liquid theory are usually couched in terms of the lifetime of single electrons `injected' above a filled Fermi sea, the lifetimes we are measuring here are those of elementary `dipoles' (particle-hole excitations).  In a Fermi liquid these relaxation rates scale the same way; in disordered systems they generally do not (see SI).  Nevertheless, in the microcanonical ensemble (relevant for optical experiments, where we do inject particles), particle hole excitations are the low-energy excitations of the system, and the marginality of their lifetime is the key diagnostic of the marginal Fermi glass\footnote{We remark, however, that the particle-hole lifetime does not distinguish between Fermi glasses and certain localized ``non-Fermi glass'' phases~\cite{nfg}.}.

To conclude, we have examined the energy relaxation rate and photon echo decay rates in P doped Si using THz 2DCS. We have discovered a host of surprising features, including relaxation rates that are linear functions of frequency with slope close to unity; echo decay rates that are temperature independent within experimental precision; energy relaxation rates that are {\it decreasing} functions of temperature over the range probed; and a doping dependence such that the relaxation rates {\it decrease} as we approach the transition to the metallic phase. We have argued that these results indicate that (i) the system is behaving as a well isolated electronic system, with coupling to phonons negligible on experimental timescales (ii) the relaxation is dominated by {\it electron-electron interactions} and arises through the interplay of strong interactions and strong disorder. By analogy with the `marginal Fermi liquid' behavior observed in strongly correlated metals, we dub the phase characterized by our experiments a `marginal Fermi glass.'

\bigskip

{\bf Methods.} Experiments were performed on nominally uncompensated phosphorous-doped silicon (Si:P) samples, which were cut from a Czochralski grown boule to a specification of \SI{5}{cm} in diameter with a P-dopant gradient along the axis. This boule was subsequently sliced and then polished down to \SI{100}{\micro\meter} wafers. Samples from this boule were previously used for studies of the THz-range conductivity in the phononless regime \cite{Helgren04a,Helgren02a} and optical pump-THz probe measurements \cite{thorsmolle2010ultrafast}.  We measured samples from 39-85$\%$ of the critical concentration of the 3D metal-insulator transition (MIT) in a regime where the localization length was of order or longer than the inter-dopant spacing.  Note that these concentrations are far higher than used in THz free electron laser (FEL) studies demonstrating photon echo at the  $1s \rightarrow 2p_0$ transition ($\sim \SI{8.29}{THz}$) \cite{greenland2010coherent,lynch2010first}.  P concentrations were calibrated with the room temperature resistivity using the Thurber scale \cite{Thurber80a}.

To perform 2D non-linear THz spectroscopy, two intense THz pulses (A and B) generated by the tilted pulse front technique and separated by a time-delay $\tau$ (Fig.~1b) are focused onto each sample in a collinear geometry (see SI for details of the experimental setup) \cite{woerner2013ultrafast,lu2017}. The transmitted THz fields were detected by standard electro-optic (EO) sampling using a \SI{30}{fs}, \SI{800}{nm} pulse that is delayed by time $t$ relative to pulse B. Displayed data was taken with a maximum electric field of \SI{50}{kV/cm} for each pulse.  A differential chopping scheme is used to extract the non-linear signal ($E_{NL}(\tau,t) = E_{AB}(\tau,t) - E_{A}(\tau,t) - E_{B}(t)$) resulting from the interaction of the two THz pulses with the sample. Here $E_{AB}$ is the transmitted signal when both THz pulses are present while $E_{A}$ and $E_{B}$ are the transmitted signals with each pulse A and B present individually. A 2D Fourier transform of $E_{NL}$ with respect to $\tau$ and $t$ gives the complex 2D spectra as a function of the frequency variables $\nu_\tau$ and $\nu_t$. Fits to Lorentzians were restricted to the central part of peaks due to the phase twisting present in these spectra as discussed in the SI.  

\let\oldaddcontentsline\addcontentsline
\renewcommand{\addcontentsline}[3]{}

 \section*{\uppercase{Acknowledgements}}
 
We would like to thank Y. Galperin, Y.-B. Kim, A. Legros, I. Martin,  A. Millis, V. Oganesyan, B. Shklovskii, and Y. Yuan for helpful discussions.  This project was supported by a now canceled DARPA DRINQS program grant. S.G. acknowledges support from NSF Grant No. DMR-1653271.
 
 \section*{\uppercase{Author contributions}}
 
FM and DC built the 2D THz setup and did experiments and analysis.  SG and RN gave theoretical support.   NPA directed the project.   All authors contributed to the writing and editing of the manuscript.

 \section*{\uppercase{Additional information}}
 
\noindent\textbf{Competing financial interests:} The authors declare no competing financial interests.\\
\bigskip

\noindent\textbf{Data availability:} All relevant data are available on reasonable request from NPA.

\let\addcontentsline\oldaddcontentsline

\normalsize

\clearpage

\setcounter{figure}{0}  
\setcounter{page}{1}
\renewcommand\thefigure{S\arabic{figure}} 
\renewcommand{\figurename}{Fig.} 

\renewcommand{\author}{}

\renewcommand{\title}{Supplementary Information: Observation of a marginal Fermi glass using THz 2D coherent spectroscopy}
\renewcommand{\affiliation}{}
\setcounter{section}{0}

\onecolumngrid

\begin{center}
	\textbf{\large \title}\\
	\medskip
	\author{Fahad Mahmood$^{1,2,3}$,  Dipanjan Chaudhuri$^1$, Sarang Gopalakrishnan$^{4,5}$,  Rahul Nandkishore$^6$, and N. P. Armitage$^1$\\
		\medskip
		\small{$^1$ \textit{Department of Physics and Astronomy, The Johns Hopkins University, Baltimore, Maryland 21218, USA}\\
			$^2$ \textit{Department of Physics, University of Illinois at Urbana-Champaign, Urbana, 61801 IL, USA}\\
			$^3$ \textit{F. Seitz Materials Research Laboratory, University of Illinois at Urbana-Champaign, Urbana, 61801 IL, USA}\\
			$^4$ \textit{Department of Engineering Science and Physics,\\CUNY College of Staten Island, Staten Island, NY 10314, USA}\\
			$^5$ \textit{Initiative for Theoretical Sciences, The Graduate Center, CUNY, New York, NY 10016, USA}\\
			$^6$ \textit{Department of Physics and Center for Theory of Quantum Matter,\\University of Colorado Boulder, Boulder, Colorado 80309, USA}\\
		}}
\end{center}

\tableofcontents

\section{Experimental setup}

To implement THz 2D coherent spectroscopy (2DCS), intense THz pulses were generated by the tilted pulse front technique by optical rectification in LiNbO$_3$ crystals \cite{hirori2011single,woerner2013ultrafast,Lu2017}. 800 nm laser pulses from an Astrella one-box Ti:Sapphire amplifier system (1 kHz, 30 fs, 7 mJ/pulse) were separated via beam splitters (BS) into two roughly identical beams, A and B. A time-delay $\tau$ was introduced between the two using a mechanical linear translation stage. Each of the laser pulses were then directed onto a diffraction grating to generate the requisite pulse front tilt for optimal phase matching. The output pulse from the grating was imaged onto the 0.6$\%$ MgO doped LiNbO$_3$ crystals to generate THz pulses. The two beams were combined using a wire-grid polarizer and then focused onto the samples in a collinear geometry with four parabolic mirrors in an 8f configuration as shown in Supplementary Fig. \ref{Schematic}. The transmitted THz fields were detected by standard electro-optic (EO) sampling in ZnTe using a third 30-fs 800 nm pulse that is delayed by time $t$ relative to pulse A. Peak electric fields on the sample are estimated to be $\sim$ 50 kV/cm for each pulse.

A differential chopping scheme is used to extract the non-linear signal defined as $E_{NL}(t,\tau) = E_{AB}(t,\tau) - E_{A}(t,\tau) - E_{B}(t)$ that results from the interaction of the two THz pulses with the sample. Here $E_{AB}$ is the transmitted signal when both THz pulses are present while $E_{A}$ and $E_{B}$ are the transmitted signals with each pulse A and B present individually. Fig. 1c-f in the main text shows $E_A$, $E_B$, $E_{AB}$, and $E_{NL}$ as a function of $t$ and $\tau$ for the 39\% sample.  A 2D Fourier transform of $E_{NL}$ with respect to $t$ and $\tau$ gave the complex 2D spectra as a function of the frequency variables $\nu_A$ and $\nu_B$. 

\begin{figure*}[h]
	\includegraphics[width=16cm]{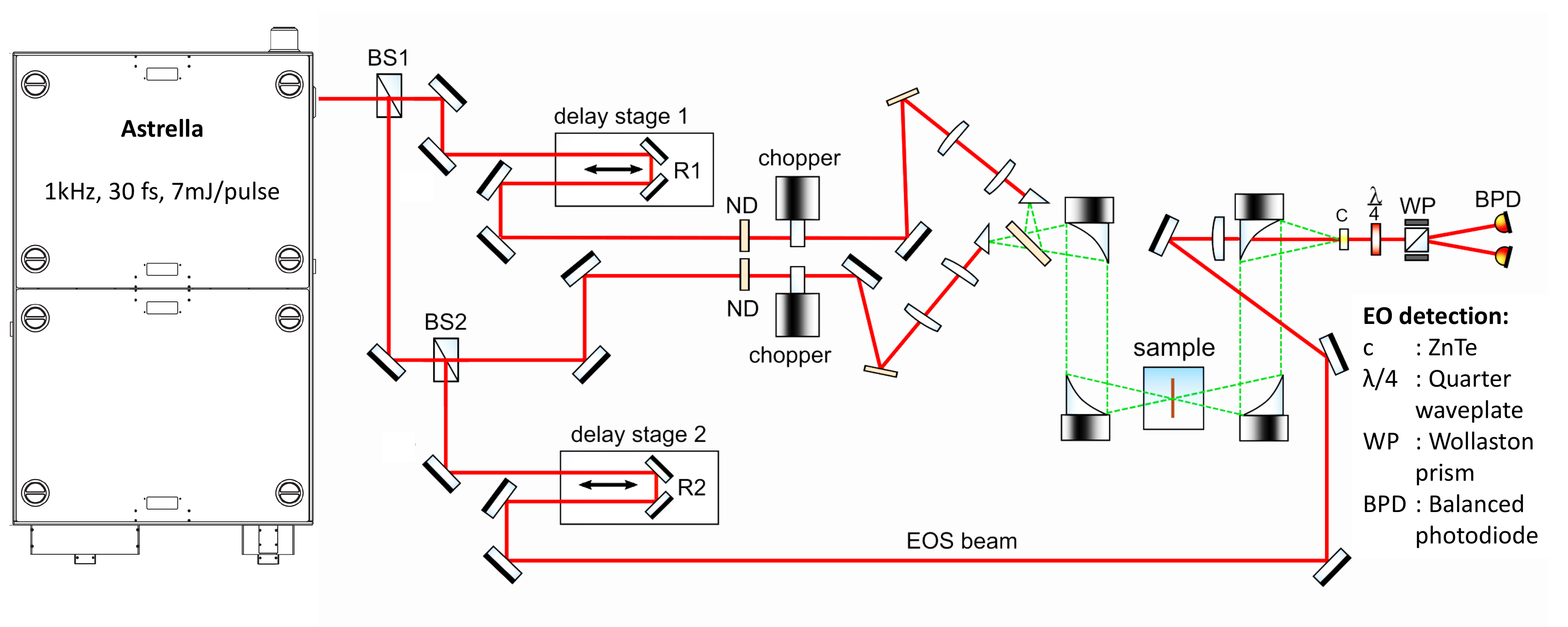}
	\caption{\label{Schematic} \textbf{Schematic diagram of the THz 2DCS setup.} The red lines denote the path of the 800 nm IR pulse and the green dashed lines show the propagation of the THz pulses.}
\end{figure*}

\section{2D nonlinear response for a single two level system}

As discussed in the main text and in Sec. \ref{dipoleham}, it is believed that one can understand the optical excitations of an electronic glass as an ensemble of random two-level systems.  Although there may be aspects of the current data that are beyond this simple picture, it is instructive to work out the theoretical expectation for the THz 2DCS response for a single idealized two-level system with finite longitudinal and transverse relaxation rates $1/T_1$ and $1/T_2$.   Related treatments exist in Refs. \cite{hamm2011concepts,Mukamel1995,hamm2005principles}.   In addition to being directly relevant to the particular physical system we are considering, this exposition should be regarded as an elementary introduction to THz 2DCS in general.

In all optical spectroscopies, one measures an emitted field in response to a time-dependent perturbing electromagnetic field.   The emitted field is caused by a time-dependent polarization and the sample response characterized by a frequency dependent complex susceptibility.   The emitted field has a 90$^\circ$ phase lag from a sample's macroscopic polarization.   In linear response this 90$^\circ$ phase lag combines with the part of the polarizability (the imaginary part) that has a 90$^\circ$ phase lag to the driving field to give an emitted field that is 180$^\circ$ out of phase with the driving field and hence destructive interference.   This gives the phenomena of absorption, which is why the imaginary part of the polarization is associated with dissipation.  This is a natural way to analyze the linear response.   However in the nonlinear regime, rather than considering a material as a system that absorbs photons, it is more illuminating to think of a time dependent polarization as a {\it source} of electromagnetic radiation.  Therefore, our task in describing the nonlinear response is to analyze the time-dependent polarization.  The time-dependent polarization is equal to the expectation value of the dipole operator e.g. $P(t) = \langle \mu(t) \rangle$. We use the ``Mukamelian" or perturbative expansion of the density matrix \cite{Mukamel1995} to model the time dependent dipole operator and eventually the nonlinear response of a two-level system.  

\subsection{Preliminaries}

The expectation value of any operator $\hat{A}$ can be expressed as trace of the density matrix multiplied by that operator.  The density matrix  $\rho   = |\Psi \rangle \langle \Psi| $ expanded in terms of basis states (typically eigenstates of $ \hat{H}_0 $ below) e.g. $|\Psi \rangle = \sum_{n} c_n |n\rangle$ is

\begin{equation}
\rho =  |\Psi \rangle \langle \Psi| =   \sum_{nm} c_n c_m^* |n\rangle \langle m|.
\end{equation}
Then for instance the expectation value for the dipole operator $ \langle \Psi(t)| \mu | \Psi(t)\rangle$ is

\begin{align}
\langle \mu \rangle & = \sum_m c^*_m \sum_n c_n   \langle m| \hat{\mu} |n \rangle \\
& = \sum_{mn}  c^*_m c_n \mu_{mn} \\
& =  \sum_{mn} \rho_{nm} \mu_{mn} =  \mathrm{Tr}(\hat{\rho} \hat{\mu}) \\
& =   \langle \hat{\rho} \hat{\mu} \rangle.
\label{PolarExpect}
\end{align}
Therefore the time dependent polarization is given by the time-dependence of the density matrix.  The time evolution of the density matrix is given by its commutator with the system Hamiltonian e.g.

\begin{equation}
\frac{d \rho}{dt }  = - \frac{i}{\hbar} [ \hat{H},\rho].
\label{Liouville}
\end{equation}
This follows from the fact that the time evolution of the density matrix is

\begin{equation}
\frac{d \rho}{dt} =  \frac{d}{dt} ( |\Psi \rangle \langle \Psi| ) =   ( \frac{d }{dt}   |\Psi \rangle)  \langle \Psi| +  |\Psi \rangle (\frac{d }{dt} \langle \Psi|  )   ).
\label{timeevolve}
\end{equation}
Using the Schroedinger equation that gives the time evolution of $|\Psi \rangle$ e.g. $ \frac{d }{dt} |\Psi \rangle =  - \frac{i}{\hbar} \hat{H} |\Psi\rangle $ (and its complex congugate) and plugging into Eq. \ref{timeevolve}, one obtains the so-called Liouville-von Neumann equation that is written as

\begin{align}
\frac{d }{dt} \rho &=   - \frac{i}{\hbar} \hat{H} |\Psi\rangle \langle \Psi | +   \frac{i}{\hbar} |\Psi\rangle \langle \Psi |  \hat{H} \\
&=  - \frac{i}{\hbar} \hat{H} \rho +   \frac{i}{\hbar} \rho \hat{H} \\
&=  - \frac{i}{\hbar} [\hat{H},\rho],
\end{align}
which is the same as Eq. \ref{Liouville} above.  The essential point is that the time dependence of $\rho$ follows from the Hamiltonian acting from both the right {\it and} left sides of the density matrix.

\subsection{Linear Response}

We will use this density matrix formalism in a number of different ways.  First let us describe two-level system's linear response.  Consider a total Hamiltonian

\begin{equation}
\hat{H}(t) = \hat{H}_0 + \hat{W}(t)
\label{FullHamiltonian}
\end{equation}
where $ \hat{H}_0$ is the time-independent part of the system Hamiltonian and  $\hat{W}(t)$ is an interaction with a time dependent field.  Here and below we assume that the set of basis states used for the density matrix $|n\rangle$ diagonalizes the system Hamiltonian  $\hat{H}_0$ as

\begin{equation}
|\Psi \rangle =  c_0 e^{-i \omega_0 t } | 0 \rangle + i c_1 e^{-i \omega_1 t } | 1 \rangle .
\end{equation}

For reasons that will be useful below, we have chosen the $c_
n$ coefficients such that a factor of $i$ is included in the definition of the wavefunction with $c_1$. In the absence of dephasing and population relaxation the elements of the density matrix are

\begin{equation}
\rho = \left(\begin{array}{cc}\rho_{00} & \rho_{01} \\ \rho_{10} & \rho_{11}\end{array}\right) = \left(\begin{array}{cc}c_0^2 & - i c_0c_1e^{i\omega_{01}t} \\ i c_0c_1e^{-i\omega_{01}t}  & c_1^2\end{array}\right),
\end{equation}
where $\omega_{01} = \omega_{1} - \omega_{0} $.   In the presence of dephasing and population relaxation standard Bloch sphere dynamics gives the time evolution of the density matrix as

\begin{equation}
\rho  = \left(\begin{array}{cc}c_0^2 + c_1^2(1 - e^{-t/T_1} )  & - i c_0c_1e^{i\omega_{01}t} e^{-t/T_2} \\ i c_0c_1e^{-i\omega_{01}t}e^{-t/T_2}  & c_1^2 e^{-t/T_1} \end{array}\right),
\label{TimeEvolve}
\end{equation}
which describes a precessing Bloch vector.  As always the homogeneous dephasing and population relaxation are related by $1/T_2 = 1/2T_1 + 1/T_2^*$ where $T_2^*$ is pure dephasing time that may be caused by fluctuations of the environment.

The interaction of the light pulse with the sample is accounted by the term $\hat{W}(t) = - \hat{\mu} E(t)$ in Eq. \ref{FullHamiltonian}.   $E(t)$ is a real valued field and the transition dipole operator is 

\begin{equation}
\hat{\mu} = \left(\begin{array}{cc} 0 & \mu_{01} \\ \mu_{10} & 0 \end{array}\right).
\end{equation}

We use a perturbative expansion of the Liouville-von Neumann equation to account for the time dependence of the density matrix under the influence of $\hat{W}(t)$.   In this analysis we use the {\it semi-impulsive} limit, where the light pulses are assumed to be short compared with any time scale of the system but long compared to the oscillation period of the light field. Therefore we describe the light field by an expression where envelopes of the pulses are approximated by $\delta$-functions, but we retain oscillations e.g.

\begin{equation}
E(t) = | \varepsilon | \delta(t)  \; \mathrm{cos}(\omega t)  = \frac{| \varepsilon |}{2}  \delta(t) (e^{i \omega t} + e^{ - i \omega t} ) 
\label{impulse}
\end{equation}
where $|\varepsilon|\delta(t)$ is the electric field envelope function and $\omega$ is the carrier frequency.  Substituting Eq. \ref{impulse} into the $\hat{W}(t)$ part of the Hamilitonian and integrating Eq. \ref{Liouville} with respect to time one gets the expression for the correction to the density matrix after one instantaneous interaction with the light pulse

\begin{equation}
\rho^{(1)} \approx \frac{i}{\hbar} \frac{ | \varepsilon|}{2} (\mu(0)\rho(-\infty) - \rho(-\infty) \mu(0) ).
\label{correction}
\end{equation}
This represents a system with a density matrix representing the ground state $\rho(-\infty)$ interacting with a light pulse at time zero.   Note that the dipole operator respresenting the light pulse operates from both the left and the right sides of the density matrix e.g. the ket and bra sides and hence that the two terms in Eq. \ref{correction} are complex congugates of each other.  Also note that in actuality {\it four} terms exist in the interaction of the light field with the ground state (the two complex conjugate terms representing the light field  multiplied by the two terms of the Liouville-von Neumann equation) however only the two in Eq. \ref{correction} are effectively non-zero as transitions can only be made to higher energies from the ground state.   This means that the $\varepsilon $ term of the electric field makes transitions only only the ket side of the density matrix whereas the $\varepsilon^* $ term makes transitions on the bra side.   Dropping the other two terms is known as the rotating wave approximation.  After interaction with the light pulse, the full density matrix has the form 

\begin{equation}
\rho(0^+) = \rho^{(0)} + \rho^{(1)} =  \left(\begin{array}{cc}1  & 0 \\ 0 & 0 \end{array}\right)   + \frac{i}{\hbar}  \frac{ | \varepsilon|}{2 }  \left(\begin{array}{cc} 0   &  -  \mu_{01}  \\    \mu_{01} & 0 \end{array}\right) .
\label{correction2}
\end{equation}

The system's density matrix then evolves freely in a manner given by Eq. \ref{TimeEvolve}, which describes an oscillating Bloch vector

\begin{equation}
\rho(t) = \rho^{(0)} + \rho^{(1)} =  \left(\begin{array}{cc}1  & 0 \\ 0 & 0 \end{array}\right)   + \frac{i}{\hbar}  \frac{ | \varepsilon|}{2 }   \left(\begin{array}{cc} 0   & - \mu_{01} e^{i\omega_{01}t}e^{-t/T_2}  \\    \mu_{01}  e^{-i\omega_{01}t}e^{-t/T_2}  & 0 \end{array}\right) .
\label{correction3}
\end{equation}

An oscillating Bloch vector will emit an electromagnetic field and so it is clear that only the time dependent part of Eq. \ref{correction3} e.g. $\rho^{(1)}$ will give an emitted wave.  The light emission is described by another interaction with the dipole operator at a time $t$ and via Eq. \ref{PolarExpect} the emitted field is given by $\langle P^{(1)}(t) \rangle =   \mathrm{Tr}(\rho(t)\mu)$.  When evaluated, this expression gives  

\begin{equation}
P^{(1)}(t) =  \varepsilon \frac{ \mu_{01}^2}{\hbar}e^{-t/T_2}  \; \mathrm{sin} (\omega_{01} t).
\label{FID}
\end{equation}

This is the so-called ``free induction decay" in the semi-impulsive limit.   When the radiation from this time-dependent polarization is added to the original incoming signal $E(t)$, destructive interference takes place and less radiation will be transmitted at a frequency $\omega_{01}$ resulting in absorption.  The Fourier transform of Eq. \ref{FID} is 

\begin{equation}
P^{(1)}(\omega) =  \varepsilon \frac{ \mu_{01}^2}{\hbar} \frac{\omega_{01} }{ \omega_{01}^2 + 1/T_2^2 - \omega^2 - i 2 \omega /T_2 }.
\label{FIDomega}
\end{equation}

As as a delta function impulse has a flat frequency profile with an electric field component $\varepsilon $ at each frequency, the Fourier transform of a response to a delta function impulse is equivalent to a response function.   So therefore the linear response defined as $P(\omega) = \chi^{(1)} E(\omega)$ (with $\varepsilon  = E(\omega) $) is 

\begin{equation}
\chi^{(1)}(\omega) =  \frac{ \mu_{01}^2}{\hbar} \frac{\omega_{01} }{ \omega_{01}^2 + 1/T_2^2 - \omega^2 - i 2 \omega /T_2 }.
\label{LinearResponse}
\end{equation}

This is the well known functional form for a Drude-Lorentz oscillator.

\subsection{Non-Linear Response}

\begin{figure*}[t]
	\includegraphics[width=17cm]{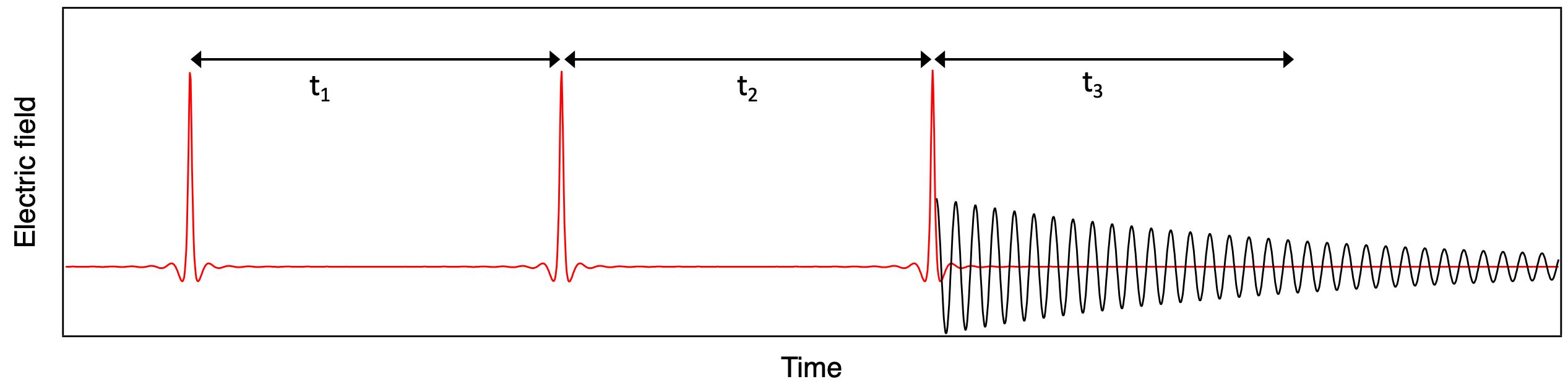}
	\caption{Generic pulse sequence for a THz 2DCS experiment.   The three red peaks represent pulses in the semi-impulsive limit.  The decaying sinusoid represents the emitted field from $\chi^{(3)}$ processes.   Note that other fields (not plotted) arising from linear response reradiation (e.g. the free-induction decay) will be emitted as a result of each single pulse by themselves, however these will be subtracted from the experimental signal by the procedure discussed in the text and do not need to be explicitly considered in the analysis. }\label{Pulses}
\end{figure*}

The above perturbative expansion of density matrices is -- frankly speaking -- a tedious method of calculating linear response.   However, it is a very powerful method to calculate the {\it nonlinear} response.   To calculate the the nonlinear response we proceed in analogous fashion starting from the time-dependent density matrix, but extending the formalism to multiple pulses.   For the third order response an analogous scheme is used e.g. semi-impulsive excitation of a ground state density matrix by the first pulse at $t=0$, free evolution of the density matrix for a time $t_1$ and then semi-impulsive excitation by the second light pulse, free evolution of the density matrix for a time $t_2$ and then semi-impulsive excitation by the third light pulse, and then emission that is described by interaction with a fourth dipole operator after an additional time $t_3$ (See Fig. \ref{Pulses}).  Again we describe the interaction of light at each step via the Liouville-von Neumman formalism using Eq. \ref{correction}, which expresses the fact that the dipole operator operates on both the ket and bra sides of the density matrix.  The third order response can be expressed compactly in terms of nested commutators as

\begin{equation}
\langle P^{(3)} \rangle = - i  \frac{| \varepsilon_2 \varepsilon_1 \varepsilon_0 |}{8 \hbar^3}  \langle \mu(t_3 + t_2 + t_1) [\mu(t_2 + t_1),[\mu(t_1),[\mu(0), \rho(-\infty)]]] \rangle.
\label{NL}
\end{equation}
Here our notation is such that $\mu(t)$ refers to action of the dipole operator at a time $t$.   It does not explicitly refer to the time dependence of the operator e.g. we are still using the Schroedinger picture for time-evolution.  The commutators can be expanded as 
\begin{align}
&\langle \mu_3[\mu_2,[\mu_1,[\mu_0,\rho(-\infty) ]]]\rangle  = \nonumber \\
& \; \; \;\; \; \; \langle \mu_3\mu_1 \rho(-\infty) \mu_0 \mu_2 \rangle - \langle \mu_2\mu_0 \rho(-\infty) \mu_1 \mu_3 \rangle  \; +  &| \; \; \;   R_1 + R_1^* \nonumber \\
& \; \; \; \; \; \; \langle \mu_3\mu_2 \rho(-\infty) \mu_0 \mu_1 \rangle - \langle \mu_1\mu_0 \rho(-\infty) \mu_2 \mu_3 \rangle  \; +  &| \; \; \;   R_2 + R_2^* \nonumber \\
&\; \; \; \; \; \; \langle \mu_3\mu_0 \rho(-\infty) \mu_1 \mu_2 \rangle - \langle \mu_2\mu_1 \rho(-\infty) \mu_0 \mu_3 \rangle  \; +  &| \; \; \;   R_4 + R_4^* \nonumber \\
& \; \; \; \; \; \; \langle \mu_3\mu_2  \mu_1 \mu_0 \rho(-\infty) \rangle - \langle \rho(-\infty)  \mu_0\mu_1  \mu_2 \mu_3 \rangle .  \;    &| \; \; \;   R_5 + R_5^*  
\label{NLexpanded}
\end{align}
The terms $R_1$, $R_1^*$, etc. denote 3rd order response functions to semi-impulsive driving fields and represent different Liouville pathways for excitation e.g. different unique sequences of manipulating the density matrices from ket and bra sides that can give emission\footnote{The terms that are missing in the notation of Ref. \cite{hamm2011concepts} $R_3$, $R_3^*$, and $R_6$, $R_6^*$ represent the double excitation of the ket and bra sides of the density matrix before de-excitation.   However, such double excitation is not allowed in a two-level system and so we do not consider these effects here.}.  Here, we have used the notation of Ref. \cite{hamm2011concepts} with regards to the dipole operators and the response functions\footnote{Note that usual definitions of the third order response function differ from our definition of the third order response by a minus sign.   Usually two factors of i that originate in the perturbative expansion of the density matrix are neglected in the definitions for third order response in 2DCS.   We have not neglected them here.}.  

We now consider the evolution of the full density matrix with first order corrections at each step as it is affected by the light pulses.   We illustrate this for $R_4$ and for notational simplicity here we ignore the changes to the density matrix during times $t_1$, $t_2$, and $t_3$. We let $\beta_n = \frac{i}{\hbar} \frac{\varepsilon_n}{2}  \mu_{01}$ and get 
\begin{align}
R_4 \; | \; \left(\begin{array}{cc} 1 & 0 \\ 0 & 0 \end{array}\right)  \xrightarrow{\mu_0 \rho}  &  \left(\begin{array}{cc} 1 & 0 \\ 0 & 0 \end{array}\right)  +   \left(\begin{array}{cc} 0 & 0 \\ \beta_0 & 0 \end{array}\right)   \xrightarrow{\rho \mu_1}  \left(\begin{array}{cc} 1 & 0 \\ \beta_0 & 0 \end{array}\right) +   \left(\begin{array}{cc} 0 & \beta_1  \\ 0 & \beta_0 \beta_1 \end{array}\right)   \xrightarrow{\rho \mu_2}   \left(\begin{array}{cc} 1 & \beta_1 \\ \beta_0 & \beta_0 \beta_1 \end{array}\right) +   \left(\begin{array}{cc} \beta_1 \beta_2 & \beta_2  \\ \beta_0 \beta_1 \beta_2  & \beta_0 \beta_2 \end{array}\right)  \xrightarrow{\mu_3 \rho } \nonumber \\  &
\left(\begin{array}{cc} 1 + \beta_1 \beta_2 & \beta_1 + \beta_2 \\ \beta_0 + \beta_0 \beta_1 \beta_2 &  \beta_0 \beta_1 + \beta_0 \beta_2 \end{array}\right) + 
\left(\begin{array}{cc} \beta_0 \beta_3 + \beta_0  \beta_1  \beta_2  \beta_3  & \beta_0 \beta_1 \beta_3 + \beta_0 \beta_2 \beta_3 \\ \beta_3 +  \beta_1 \beta_2 \beta_3  & \beta_1 \beta_3 + \beta_2 \beta_3  \end{array}\right).
\label{FullDM}
\end{align}
Note that only the diagonal elements of the second term on the bottom line of Eq. \ref{FullDM} contain dependencies on the radiation dipole operator $\beta_3$ that gives an external field.  The trace of this matrix gives a quantity proportional to the reradiated field.  If one applies the protocol considered in the text and discussed above where one analyzes only the {\it nonlinear} signal defined as $E_{NL} = E_{AB}$ - $E_{A}$ - $E_{B}$ this subtracts off the diagonal parts of the density matrix (e.g. $\beta_1 \beta_3$) that gives the linear response contributions to the reradiating field.   The residual contains only effects of the $\chi^{(3)}$ nonlinear susceptibility.   Therefore going forward, we can analyze the $\chi^{(3)}$ response by considering only the repeated operation of the dipole operator on the {\it corrections} to the density matrices after each field interaction.

Here we illustrate the evolution of the density matrix for the response function $R_1$, by now considering the full time dependencies, but at each step of interaction with the light pulse or time evolution retaining only the first order correction to the density matrix at each step.   This gives

\begin{align}
R_1 \; | \; &\left(\begin{array}{cc} 1 & 0 \\ 0 & 0 \end{array}\right)  \xrightarrow{\rho(t)  \mu_{0} } \frac{ |\varepsilon_0|}{2} \left(\begin{array}{cc} 0 & i \frac{ \mu_{10} }{\hbar} \\ 0 & 0 \end{array}\right) \xrightarrow{ t_1} \frac{ |\varepsilon_0|}{2} \left(\begin{array}{cc} 0 & i \frac{ \mu_{10} }{\hbar} e^{i \omega_{01} t_1  } e^{-t_1/T_2} \\ 0 & 0 \end{array}\right)   \xrightarrow{\mu_1 \rho(t) }  \frac{|\varepsilon_0  \varepsilon_1|}{4} \left(\begin{array}{cc} 0 & 0\\ 0 & -  \frac{ \mu_{10}^2 }{\hbar^2} e^{i \omega_{01} t_1  } e^{-t_1/T_2}   \end{array}\right)   \xrightarrow{t_2}\nonumber \\  & \;\;\;\;\;\;\;\;\;\; \frac{|\varepsilon_0  \varepsilon_1|}{4} \left(\begin{array}{cc} 0 & 0\\ 0 & -  \frac{ \mu_{10}^2 }{\hbar^2} e^{i \omega_{01} t_1  } e^{-t_1/T_2} e^{-t_2/T_1}   \end{array}\right)  \xrightarrow{\rho(t) \mu_2} \frac{|  \varepsilon_0  \varepsilon_1  \varepsilon_2 |}{8} \left(\begin{array}{cc} 0 & 0\\  - i  \frac{ \mu_{10}^3 }{\hbar^3} e^{i \omega_{01} t_1  } e^{-t_1/T_2} e^{-t_2/T_1}  & 0  \end{array}\right)  \xrightarrow{t_3} \nonumber \\ & \;\; \frac{|  \varepsilon_0  \varepsilon_1  \varepsilon_2 |}{8}  \left(\begin{array}{cc} 0 & 0\\  - i  \frac{ \mu_{10}^3 }{\hbar^3} e^{i \omega_{01} (t_1 - t_3 ) } e^{-(t_1 + t_3)/T_2} e^{-t_2/T_1}  & 0  \end{array}\right)  \xrightarrow{\mu_3 \rho(t)} \frac{|  \varepsilon_0  \varepsilon_1  \varepsilon_2 |}{8} \left(\begin{array}{cc} - i  \frac{ \mu_{10}^4 }{\hbar^3} e^{i \omega_{01}( t_1 -  t_3 )  } e^{-(t_1 + t_3)/T_2} e^{-t_2/T_1} & 0 \\ 0 & 0 \end{array}\right).
\end{align}
Here $\mu_n$ represents the action of the electric field pulse at the time $t_n$.  When combined with the complex conjugated term $R_1^*$ one gets the $R_1 + R_1^*$ contribution to the total response in the semi-impulsive limit as

\begin{equation}
P^{(3)}_{R_1 + R_1^*}(t_1,t_2,t_3) =  \frac{|  \varepsilon_0  \varepsilon_1  \varepsilon_2 |}{4} \frac{ \mu_{10}^4 }{\hbar^3}  e^{-(t_1 + t_3)/T_2} e^{-t_2/T_1} \mathrm{sin}\; [ \omega_{01}( t_1 -  t_3 )  ].\label{R1}
\end{equation}

The other terms in the expansion of the density matrix can be evaluated in a similar fashion by considering each term in Eq. \ref{NLexpanded}.   In so doing, one finds  that $P^{(3)}_{R_1 + R_1^*}  = P^{(3)}_{R_2 + R_2^*} $ and that $P^{(3)}_{R_4 + R_4^*}  = P^{(3)}_{R_5 + R_5^*} $.   The response for $R_4 + R_4^*$ is

\begin{equation}
P^{(3)}_{R_4 + R_4^*}(t_1,t_2,t_3) = - \frac{|  \varepsilon_0  \varepsilon_1  \varepsilon_2 |}{4} \frac{ \mu_{10}^4 }{\hbar^3}  e^{-(t_1 + t_3)/T_2} e^{-t_2/T_1} \mathrm{sin}\; [ \omega_{01}( t_1 +  t_3 )  ]. \label{R4}
\end{equation}

Note the different time dependencies of the oscillating term on $t_1$ and $t_3$ in Eqs. \ref{R1} and \ref{R4}.  The coherent time evolution of the state is in opposite directions for times $t_1$ and $t_3$ for $R_1 + R_1^*$, whereas they are in the same direction for $R_4 + R_4^*$.   In the former case, this is known as rephasing and gives the phenomena of photon echo for response $R_1 + R_1^*$ (and  $R_2 + R_2^*$) making it the most important spectroscopic contribution.  $R_4 + R_4^*$ and $R_5 + R_5^*$ are known as non-rephasing contributions.

In an actual experiment one does not typically use three pulses.   Instead two pulses (A and B) are used and then with the scheme of Fig. \ref{Pulses} either $t_1 \rightarrow 0$ or $t_2 \rightarrow 0 $.  In the below, we assume that pulse A arrives at a time zero, pulse B arrives at a later time $\tau$, and the electric fields are measured at time $t$ that is measured in reference to pulse B.  We refer to this as the AB pulse sequence.  One has then four separate contributions that corresponds rephasing and non-rephasing versions of the first pulse giving two field interactions or the second pulse giving two field interactions.  They are

\begin{align}
P^{(3)}_{R_1 + R_1^*}(\tau,t:AB) =&\; \; \; \;  \frac{|  \varepsilon_A^2 \varepsilon_B |}{4}  \frac{ \mu_{10}^4 }{\hbar^3}  e^{-\tau/T_1}e^{-t/T_2} \mathrm{sin}\; [ \omega_{01}(  -  t )  ]  &t_1 \rightarrow 0 & \; \; \; \; \; \; \mathrm{AB \; Pump-probe} \label{ExperimentalSignal1}  \\ 
P^{(3)}_{R_4 + R_4^*}(\tau,t:AB) =& - \frac{|  \varepsilon_A^2 \varepsilon_B |}{4}  \frac{ \mu_{10}^4 }{\hbar^3}   e^{-\tau/T_1}e^{-t/T_2} \mathrm{sin}\; [ \omega_{01}(   t )  ]  &t_1 \rightarrow 0  & \; \; \; \; \; \; \mathrm{AB \; Pump-probe} \label{ExperimentalSignal2}\\
P^{(3)}_{R_1 + R_1^*}(\tau,t:AB) =& \; \; \; \;  \frac{|  \varepsilon_A  \varepsilon_B^2  |}{4} \frac{ \mu_{10}^4 }{\hbar^3}  e^{-(\tau + t)/T_2}  \mathrm{sin}\; [ \omega_{01}( \tau -  t )  ]  &t_2 \rightarrow 0  & \; \; \; \; \; \; \mathrm{AB \; Rephasing} \label{ExperimentalSignal3}\\
P^{(3)}_{R_4 + R_4^*}(\tau,t:AB) =& -   \frac{|  \varepsilon_A  \varepsilon_B^2  |}{4}  \frac{ \mu_{10}^4 }{\hbar^3}  e^{-(\tau +t)/T_2} \mathrm{sin}\; [ \omega_{01}( \tau +  t )  ] .  &t_2 \rightarrow 0  & \; \; \; \; \; \; \mathrm{AB \; Non-rephasing}
\label{ExperimentalSignal4}
\end{align}

Note $t \geq 0$ and $\tau \geq 0$ for the equations above and by causality, $P^{(3)} = 0$ for $t < 0$ or $\tau < 0$.  Here AB refers to the pulse sequence where pulse A arrives before pulse B. One can see that for $t_1 \rightarrow 0$, the $R_1 + R_1^*$ and $R_4 + R_4^*$ contributions become identical.  This contribution which depends only on $\omega_{01}$ and $1/T_1$ is referred to as the pump-probe (PP) signal.  The pump-probe $R_1$ contribution corresponds to a bra side action (e.g. operation from the right) of $\varepsilon^*_A$ on the density matrix, then immediate action of $\varepsilon_A$ on the ket side, then time evolution for a time $\tau$, then the action of $\varepsilon_B$ on the bra side of $\rho$, and then emission at a time $t$.  This gives a response proportional to $(\varepsilon^*_A \varepsilon_A) \varepsilon_B$.  The pump-probe $R_1^*$ contribution gives a ket side interaction with $\varepsilon_A$, and then immediate action of $\varepsilon^*_A$ on the bra side, then time evolution for a time $\tau$, then the action of $\varepsilon^*$ on the ket side of $\rho$, and then emission at a time $t$.  This gives a response proportional to ($\varepsilon_A \varepsilon^*_A) \varepsilon^*_B$.  The pump-probe $R_4$ corresponds to a ket side interaction with $\varepsilon_A$, an immediate action of $\varepsilon^*_A$ on the bra side, the time evolution for a time $\tau$, then (in contrast to $R_1^*$) the action of $\varepsilon^*$ on the {\it bra} side of $\rho$, and then emission at a time $t$.   This gives a response proportional to $(\varepsilon_A \varepsilon^*_A) \varepsilon_B$, which will be the same as $R_1$.

\begin{figure*}[t]
	\includegraphics[width=11cm]{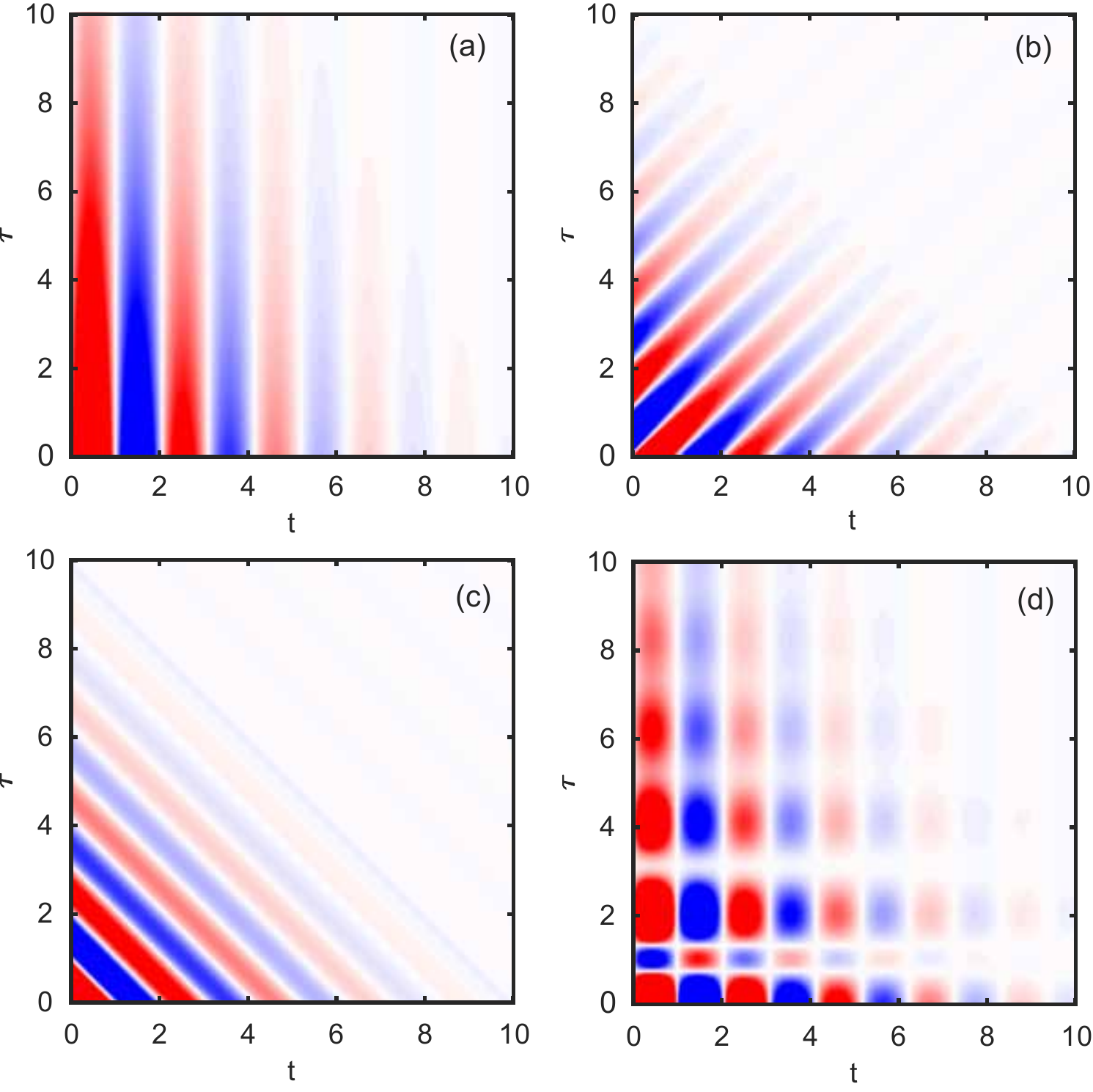}
	\caption{Plots of the 2D maps of the time dependencies of the expressions in Eqs. \ref{ExperimentalSignal1}  - \ref{ExperimentalSignal4} as a function of $\tau$ (vertical) and $t$ (horizontal).  For these plots $T_1 = 4$, $T_2 = 2$, and $\omega_{01} = 3$ was used.  (a) The pump-probe signal from Eqs. \ref{ExperimentalSignal1} and \ref{ExperimentalSignal2}.  (b) The rephasing signal (R) from Eq. \ref{ExperimentalSignal2}. (c) The non-rephasing signal (NR) from Eq. \ref{ExperimentalSignal2}.  (d) The sum of all 3 experimental signals.   This is what is actually measured in an experiment. }
	\label{TimeTraces2D}
\end{figure*}

\begin{figure*}[h]
	\includegraphics[width=16cm]{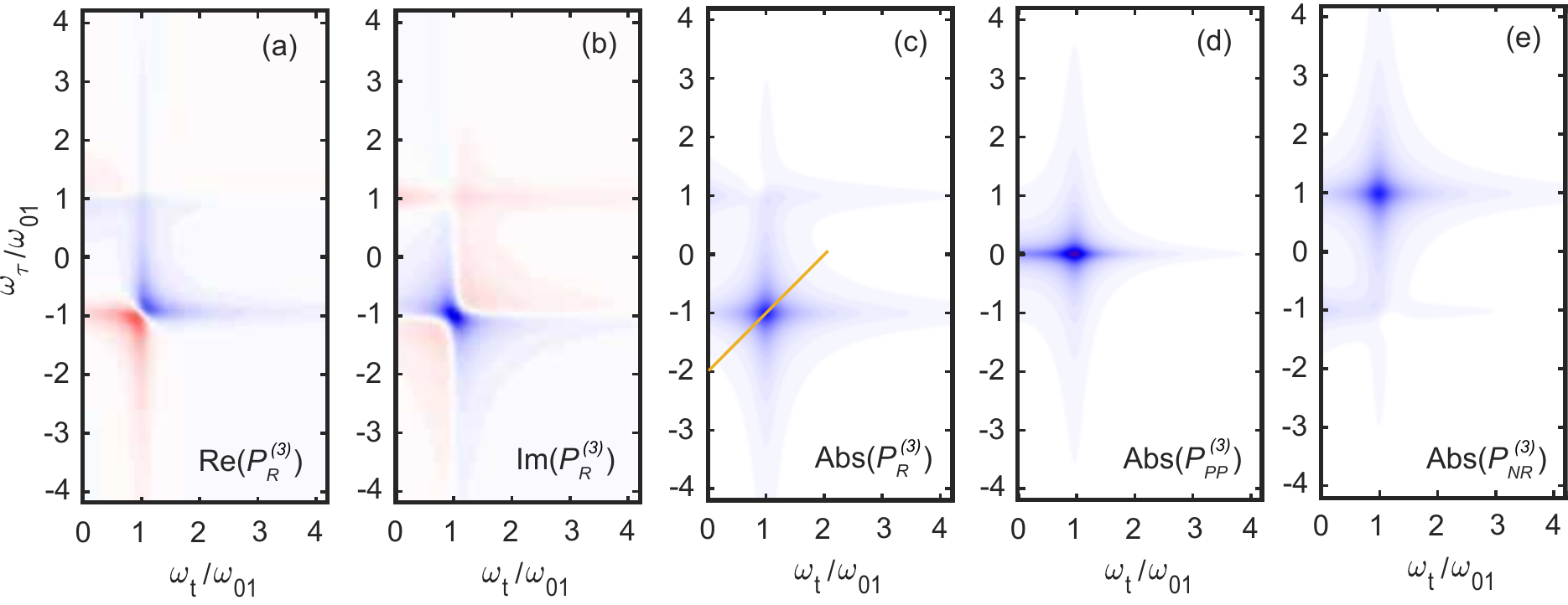}
	\caption{Fourier transform of the 2D data for the AB sequence when pulse A precedes pulse B (a) Plot of the 2D Fourier transform of the real part of the R signal (b) imaginary part of the R signal (c) Absolute value of the R signal (d) Absolute value of the PP signal (e) Absolute value of the NR signal. The plots correspond to the expressions in Eq. \ref{ExperimentalSignalFT2}-\ref{ExperimentalSignalFT4}. The faint ``ghost crosses" come from the resonances which are found at $\omega_t < 0$.    Spectra are symmetric around the origin.}
	\label{FT_2D}
\end{figure*}

\begin{figure*}[t]
	\includegraphics[width=11cm]{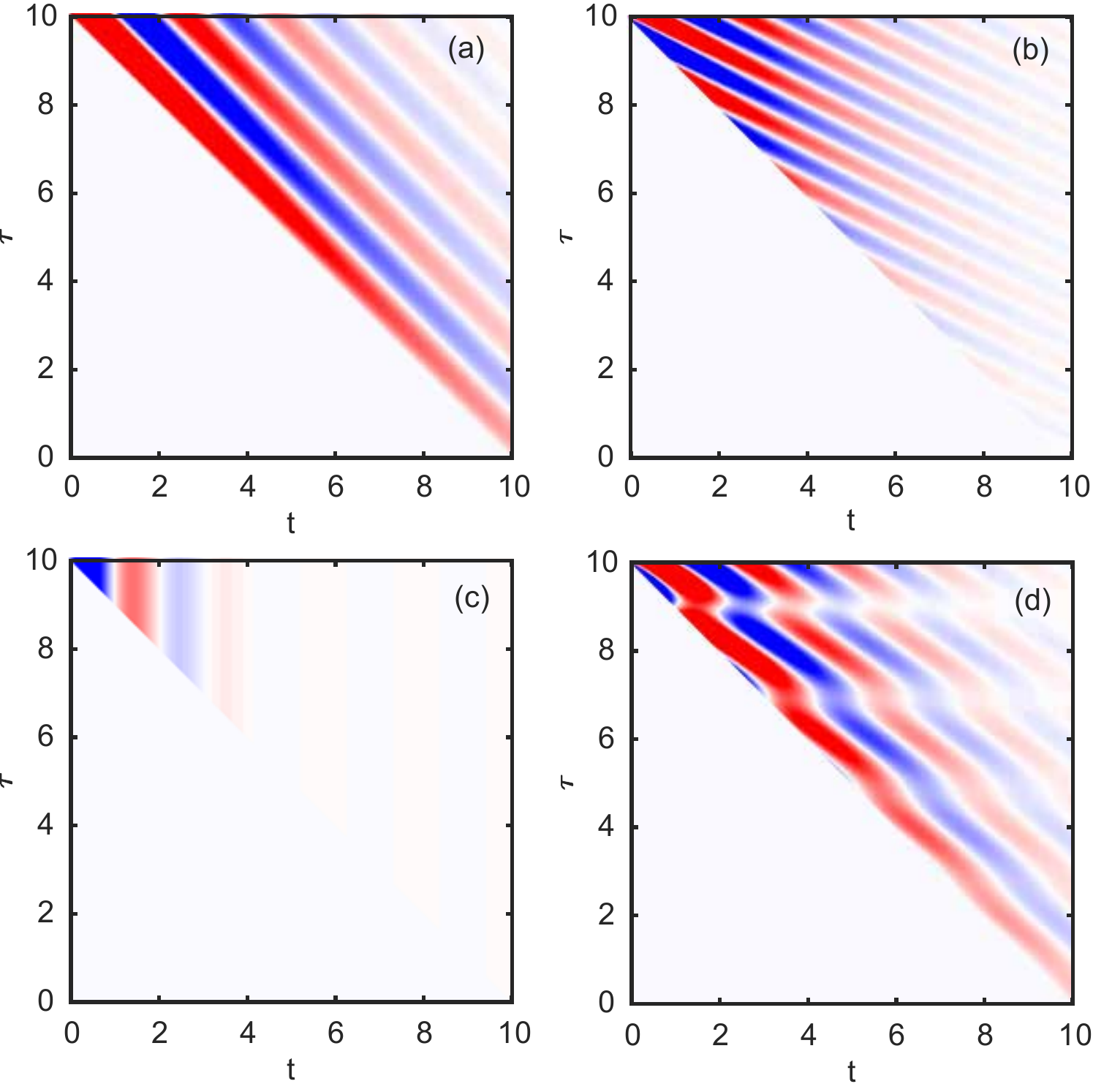}
	\caption{Plots of the 2D maps of the time dependencies of the expressions in Eqs. \ref{ExperimentalSignal2BA}  - \ref{ExperimentalSignal4BA} for the BA pulse sequence as a function of $\tau$ (vertical) and $t$ (horizontal).  For these plots again $T_1 = 4$, $T_2 = 2$, and $\omega_{01} = 3$ was used.  (a) The pump-probe signal from Eqs. \ref{ExperimentalSignal2BA}.  (b) The rephasing signal (R) from Eq. \ref{ExperimentalSignal3BA}.  (c) The non-rephasing signal (NR) from Eq. \ref{ExperimentalSignal4BA}.  (d) The sum of all 3 experimental signals.}
	\label{TimeTraces2D_BA}
\end{figure*}

For $t_2 \rightarrow 0$, there are two distinct contributions.   These are known as the rephasing (R) (Eq. \ref{ExperimentalSignal3}) and non-rephasing (NR) (Eq. \ref{ExperimentalSignal4}) signals.   The rephasing $R_1$ contribution corresponds to a bra side action of $\varepsilon^*_A$ on the density matrix, then time evolution for a time $\tau$, then action of $\varepsilon_B$ on the ket side, then the immediate action of $\varepsilon_B$ on the bra side of $\rho$, and then emission at a time $t$.  This gives a response proportional to $\varepsilon^*_A (\varepsilon_B \varepsilon_B)$. The rephasing $R_1^*$ contribution corresponds to a ket side action of $\varepsilon_A$ on the density matrix, then time evolution for a time $\tau$, then action of $\varepsilon^*_B$ on the bra side, then the immediate action of $\varepsilon^*_B$ on the ket side of $\rho$, and then emission at a time $t$.  This gives a response that depends on $\varepsilon_A (\varepsilon^*_B \varepsilon^*_B)$.  The non-rephasing $R_4$ contribution corresponds to a ket side action of $\varepsilon_A$ on the density matrix, then time evolution for a time $\tau$, then action of $\varepsilon^*_B$ on the bra side, then the immediate action of $\varepsilon_B$ on the bra side of $\rho$, and then emission at a time $t$, which gives a response that depends on $\varepsilon_A (\varepsilon^*_B \varepsilon_B)$.   The non-rephasing contribution is also known as the perturbed free-induction decay \cite{kuehn2011two} as it can be see as the perturbation by pumping of a free-induction decay.  These rather complicated dependencies and different Liouville pathways can be represented compactly in terms of ``Feynman diagrams".  We refer the interested reader to the literature \cite{hamm2011concepts} for their use and interpretation.

We plot these function and their sum in Fig. \ref{TimeTraces2D}.  Note that the different contributions have different characteristics with regards to the directions that oscillation occurs and that the signal decays.  This is the essential utility of 2DCS.  For the AB sequence the pump probe signals oscillate and shows $T_2$ decay in the $\hat{t}$ directions, but $T_1$ decay in the $\hat{\tau}$ direction.   The rephasing contribution shows oscillations in the $(\hat{\tau} - \hat{t})/\sqrt{2}$ direction, but $T_2$ decay in the orthogonal $(\hat{\tau} + \hat{t})/\sqrt{2}$ direction.   The non-rephasing contribution shows both its oscillations and $T_2$ decay in the $(\hat{\tau}  + \hat{t})/\sqrt{2}$ direction.  These differences are essential when considering the role of inhomogeneous broadening on the experimental signal and limits the usefulness of the NR signal (see discussion below) when multiple closely spaced oscillators exist.

\begin{figure*}[h]
	\includegraphics[width=11cm]{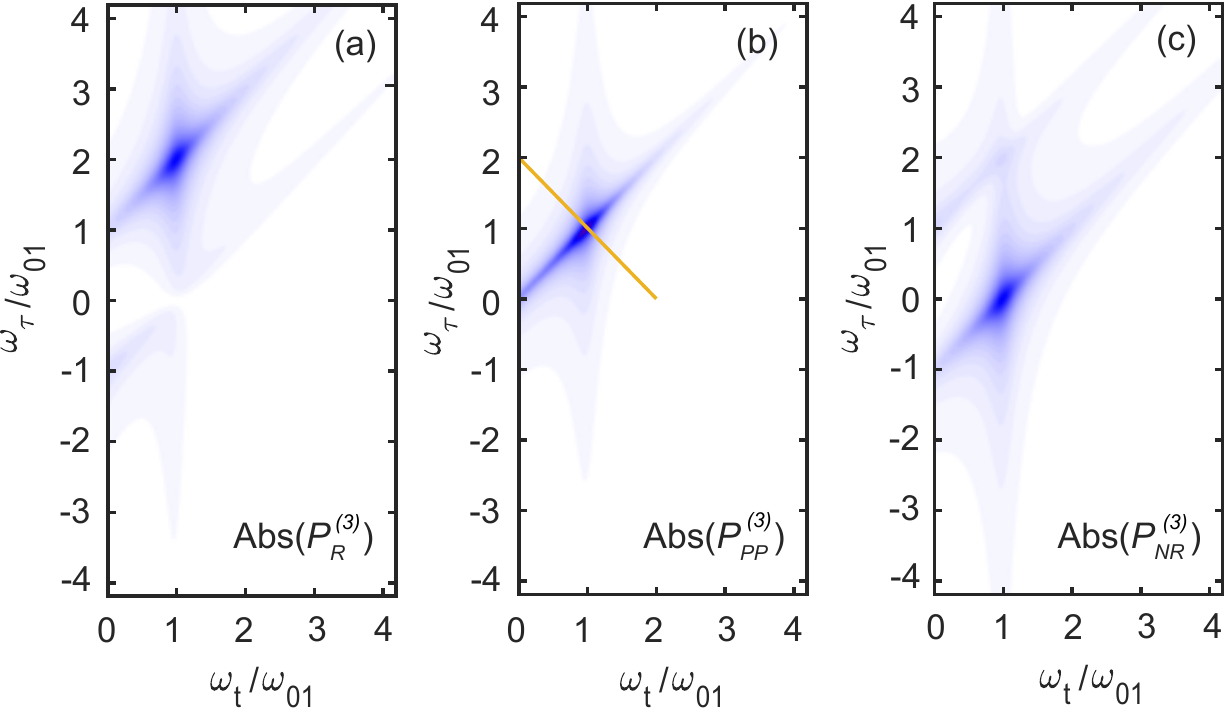}
	\caption{Fourier transforms for the BA pulse sequence (a) Absolute value of the R signal (d) Absolute value of the PP signal (e) Absolute value of the NR signal. The plots correspond to the expressions in Eqs. \ref{ExperimentalSignal2BA}-\ref{ExperimentalSignal4BA}.}
	\label{FT_2D_BA}
\end{figure*}

In the analysis of experimental data one typically takes the Fourier transforms of the experimental quantities.   The Fourier transforms of the full PP and R signals as a function of $\tau$ and $t$ in the AB pulse sequence are 

\begin{align}
P^{(3)}_{PP}(\omega_\tau,\omega_t:AB) =& 
\frac{i}{2} |  \varepsilon_A^2  \varepsilon_B |  \frac{ \mu_{10}^4 }{\hbar^3}  \frac{ 1 }{  1/T_1 + i \omega_\tau  }  \;  \Big[ \frac{1}{1/T_2 + i (\omega_t - \omega_{01})}  - \frac{1}{1/T_2 + i (\omega_t + \omega_{01})}  \Big] \nonumber \\
=& - |  \varepsilon_A^2  \varepsilon_B |  \frac{ \mu_{10}^4 }{\hbar^3}  \frac{ 1 }{  1/T_1 + i \omega_\tau  }  \;  \Big[ \frac{\omega_{01} }{ \omega_{01}^2 + 1/T_2^2 - \omega_t^2 - i 2 \omega_t /T_2 }  \Big]
\label{ExperimentalSignalFT2},\\
P^{(3)}_{R}(\omega_\tau,\omega_t:AB) =&  \frac{-i}{4} |  \varepsilon_A  \varepsilon_B^2 | \frac{ \mu_{10}^4 }{\hbar^3}  \Big[  \frac{1}{1/T_2 + i (\omega_\tau + \omega_{01})}\frac{1}{1/T_2 + i (\omega_t - \omega_{01})}  -   \frac{1}{1/T_2 + i (\omega_\tau - \omega_{01})} \frac{1}{1/T_2 + i (\omega_t + \omega_{01})} \Big]  .
\label{ExperimentalSignalFT4}
\end{align}
We plot these functions in Fig. \ref{FT_2D}.

In the response functions above peaks are found for the PP signal in the AB pulse sequence at ($\omega_\tau = 0, \omega_t = \pm \omega_t$) with a $\hat{\omega}_\tau$ direction width set by $1/T_1$ and a $\hat{\omega}_t$ direction width set by $1/T_2$.   The R signal is found at ($\omega_\tau = \mp \omega_{01}, \omega_t = \pm \omega_{01}$) with a width that is set by $1/T_2$ in the $(\hat{\omega}_\tau + \hat{\omega}_t)/\sqrt{2}$ direction and is infinitesimally narrow in the orthogonal direction.   The NR signal is found at ($\omega_\tau = \pm \omega_{01}, \omega_t = \pm \omega_{01}$) and again is broadened in the $(\hat{\omega}_\tau + \hat{\omega}_t)/\sqrt{2}$ direction by $1/T_2$ and is not broadened at all in the orthogonal direction.  Note that an intrinsic feature of 2DCS spectra in general and that can be seen explicitly in Eqs. \ref{ExperimentalSignalFT2} and \ref{ExperimentalSignalFT4} is ``phase twisting." Each frequency axis gives a complex contribution to the response and the overall 2D response is the product of these complex contributions.  Therefore as shown in Fig. \ref{FT_2D}, neither the real or imaginary parts of the $\chi^{(3)}$ response correspond to purely absorptive spectra characterized by simple peaked lineshapes.    They show more complicated mixed absorptive and dispersive character, which complicates their analysis.  A phasing procedure used in previous THz 2DCS experiments \cite{kuehn2011two} to get purely absorptive lineshapes is challenging to implement here, because of the extreme inhomogeneous broadening and overlapping contributions in 2D frequency space.  For this reason we chose to analyze the magnitudes of the nonlinear response in the experimental data which can be approximated with a Lorentzian lineshape to extract the widths $1/T_1$ and $1/T_2$. This is discussed in detail below.   

In the treatment above, pulse A precedes pulse B. However in our experiment we scan pulse A through pulse B, such that we can acquire data where pulse B precedes pulse A.   One can get the contribution of these to the experiment by the substitutions into Eqs. \ref{ExperimentalSignal1} - \ref{ExperimentalSignal4} of $t \rightarrow t + \tau$ and $\tau \rightarrow - \tau$.   These time dependencies (plotted in Fig. \ref{TimeTraces2D_BA}) are

\begin{align}
P^{(3)}_{R_4 + R_4^*}(-\tau,t+\tau:BA) =& - \frac{|   \varepsilon_B^2 \varepsilon_A |}{4}  \frac{ \mu_{10}^4 }{\hbar^3}   e^{\tau/T_1}e^{-(t + \tau)/T_2} \mathrm{sin}\; [ \omega_{01}(   t + \tau )  ]  &  & \; \; \; \; \; \; \mathrm{BA\;Pump-probe} \label{ExperimentalSignal2BA}\\
P^{(3)}_{R_1 + R_1^*}(-\tau,t+\tau:BA) =& \; \; \; \;  \frac{|  \varepsilon_B \varepsilon_A^2    |}{4} \frac{ \mu_{10}^4 }{\hbar^3}  e^{- t/T_2}  \mathrm{sin}\; [ \omega_{01}( -2\tau -  t )  ]  &  & \; \; \; \; \; \; \mathrm{BA\;Rephasing} \label{ExperimentalSignal3BA}\\
P^{(3)}_{R_4 + R_4^*}(-\tau,t+\tau:BA) =& -   \frac{| \varepsilon_B \varepsilon_A^2   |}{4}  \frac{ \mu_{10}^4 }{\hbar^3}  e^{-t/T_2} \mathrm{sin}\; [ \omega_{01}( t )  ] .  &  & \; \; \; \; \; \; \mathrm{BA \;Non-rephasing}
\label{ExperimentalSignal4BA}
\end{align}

Note $t \geq 0$ and $\tau \leq 0$ for the equations above and by causality, $P^{(3)} = 0$ for $t+\tau < 0$.  The magnitude of Fourier transform of these functions are plotted in Fig. \ref{FT_2D_BA}.  Of course there is nothing fundamentally different in the BA sequence than the AB sequence.   The differences are only in the labels of the timing of arrival of peaks and measurement.

\begin{figure*}[t]
	\includegraphics[width=8cm]{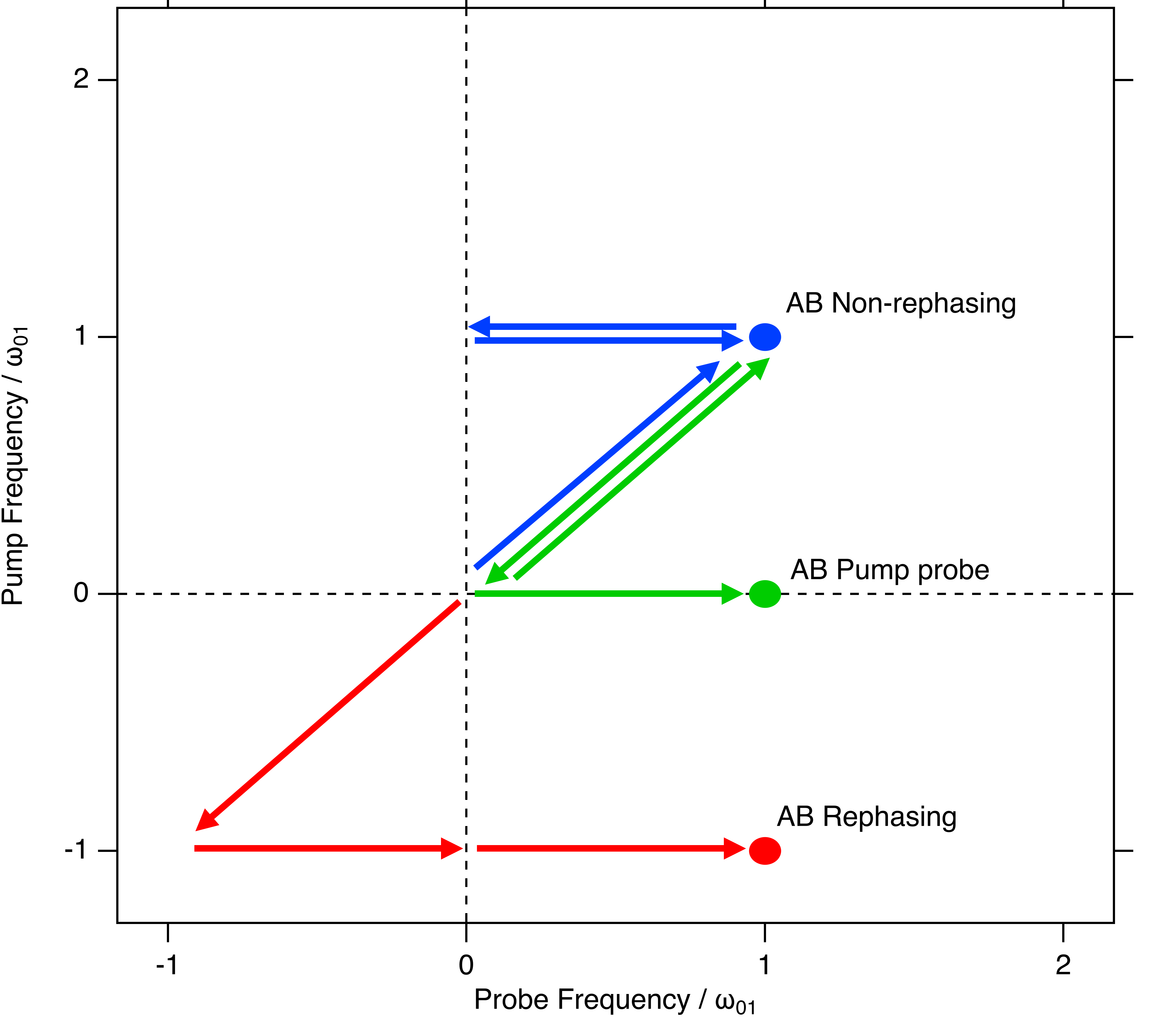}
	\includegraphics[width=8cm]{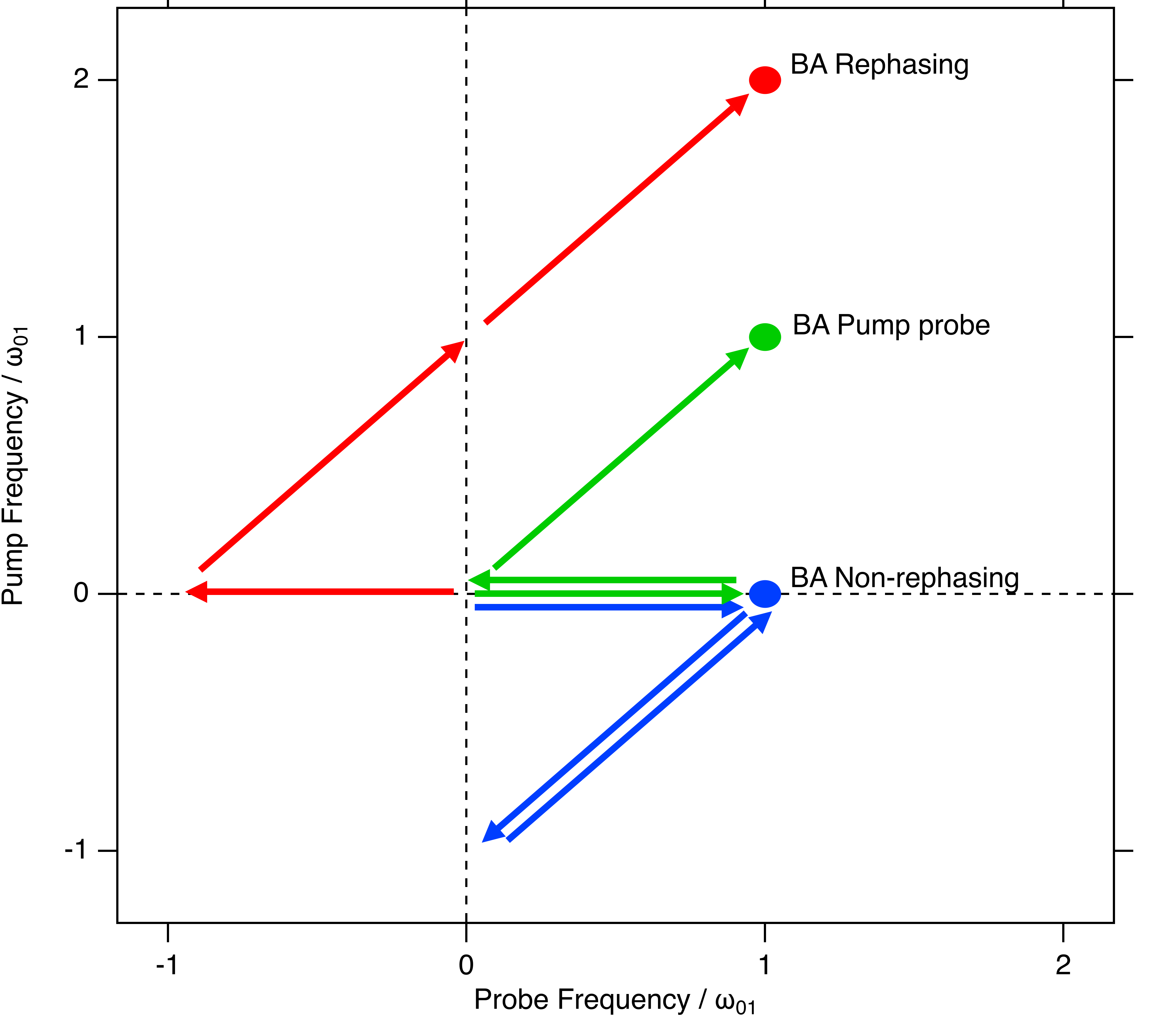}
	\caption{Frequency vector scheme for the AB left) and BA right) pulse sequences. $ \vec{\omega}_A= \omega_{01}(\hat{\omega}_\tau + \hat{\omega}_t)$ is a diagonal vector and $\vec{\omega}_B= \omega_{01} \hat{\omega}_t$ is a horizontal vector.  Conjugated electric field pulses are represented by a reversed vector.  Experimental spectra for our experimental scheme is a superposition of the two pulse sequences. }
	\label{FreqVectors}
\end{figure*}

Identification of the various contributions to the nonlinear response is made easier by the realization that a scheme of ``frequency vectors" can be used to identify the particular contribution to the response \cite{kuehn2011two,woerner2013ultrafast}.  Given a weakly non-linear oscillator with frequency $\omega_{01}$, the frequency vectors that correspond to pulses A and B are $\vec{\omega}_A = \omega_{01}(\hat{\omega}_\tau + \hat{\omega}_t)$ and $\vec{\omega}_B = \omega_{01} \hat{\omega}_t$.  Conjugated electric field pulses are represented by a reversed vector. Thus the PP contribution in the AB sequence can be placed by $\vec{\omega}_{PP:AB} =  \vec{\omega}_A - \vec{\omega}_A +  \vec{\omega}_B$.   The R signal with AB pulses is $\vec{\omega}_{R:AB} =  - \vec{\omega}_A 
+\vec{\omega}_B +  \vec{\omega}_B$ and the NR signal with AB pulses is $\vec{\omega}_{NR:AB} =   \vec{\omega}_A 
-\vec{\omega}_B +  \vec{\omega}_B$.  The PP contribution in the BA sequence is $\vec{\omega}_{PP:BA} =  \vec{\omega}_B - \vec{\omega}_B +  \vec{\omega}_A$.  The R contribution in the BA sequence is $\vec{\omega}_{R:BA} =  -\vec{\omega}_B + \vec{\omega}_A +  \vec{\omega}_A$. The NR contribution in the BA sequence is $\vec{\omega}_{NR:BA} =  \vec{\omega}_B - \vec{\omega}_A +  \vec{\omega}_A$.   These contributions and the resulting frequency vector scheme can be seen in Fig. \ref{FreqVectors}.

The special importance of the PP and R contributions can be seen when one considers the response of a physical system that has many closely spaced overlapping resonances.  The present case of an electron glass is clear realization of such a scenario.   If the different excitations are largely independent (e.g. non-coupled), then we can understand the total system response as a convolution of the response of a single two-level system with the density of states of two-level excitations.  For the simple case of a normal distribution of two-level systems, one gets for the convolution of the impulse responses given in Eqs. \ref{R1} and \ref{R4} as

\begin{figure*}[t]
	\includegraphics[width=11cm]{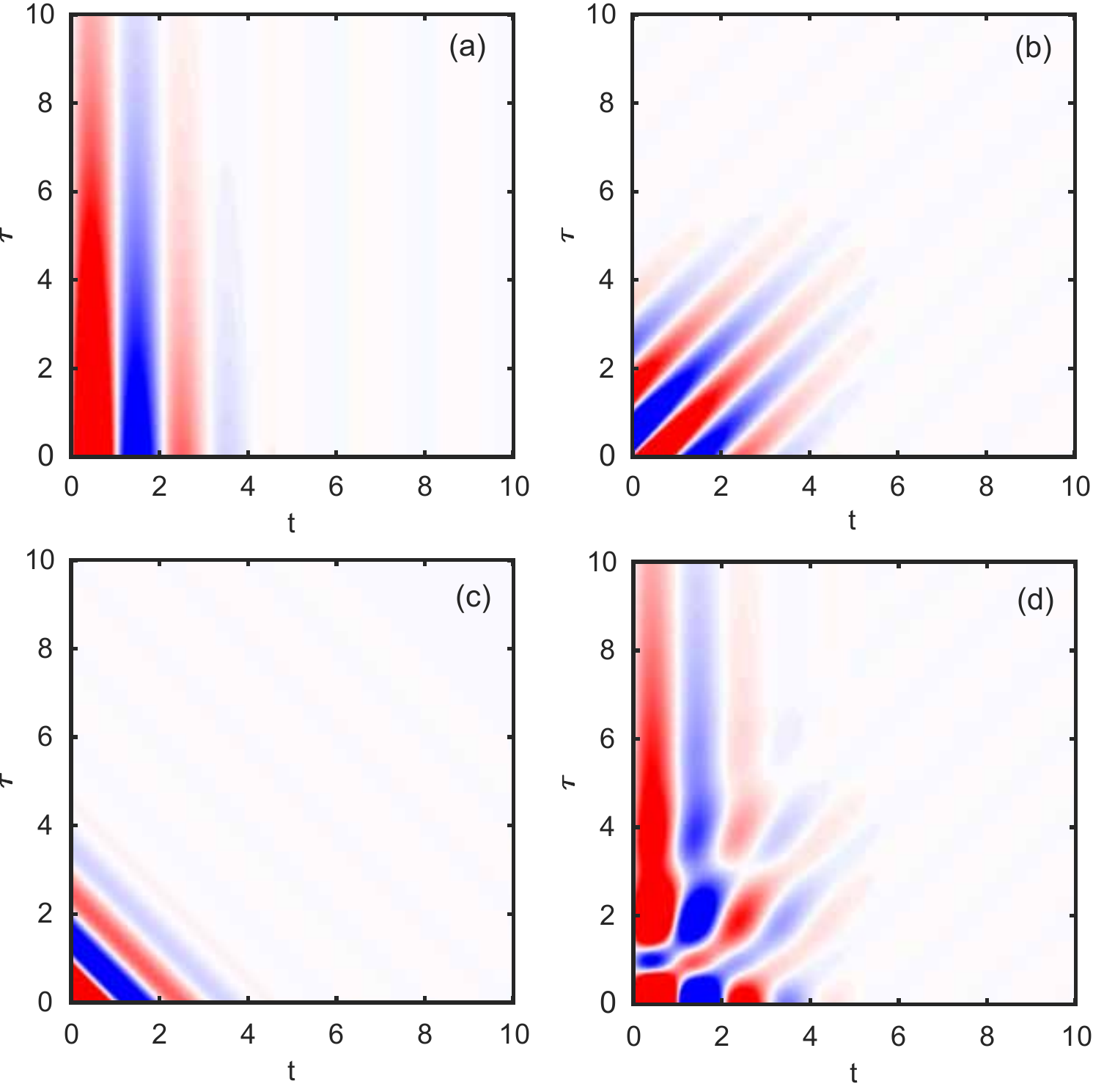}
	\caption{Plots of the 2D maps of the time dependencies of the response contributions using $\overline{ \omega}_{01}  =  3 $ and a distribution of resonance frequencies $\sigma = 0.5$.  Again $T_1 = 4$ and $T_2 = 2$.  (a) The pump-probe (PP) signal.  (b) The rephasing signal (R).  (c) The non-rephasing signal (NR).  (D) The sum of all 3 experimental signals. }
	\label{TimeTracesCV2D}
\end{figure*}

\begin{align}
\overline{P^{(3)}_{R_1 + R_1^*} } &=  \frac{|  \varepsilon_0  \varepsilon_1  \varepsilon_2 |}{4} \frac{ \mu_{10}^4 }{\hbar^3}   e^{-(t_1 + t_3)/T_2} e^{-t_2/T_1}  \int d\omega_{01} \Big( \frac{e ^{ - {(\omega_{01} -   
			\overline{ \omega}_{01}  )^2}  / 2 \sigma^2}}{\sigma \sqrt{2 \pi}} \Big) \;  \mathrm{sin}\; [ \omega_{01}( t_1 -  t_3 )  ]  \nonumber \\ &=  \frac{|  \varepsilon_0  \varepsilon_1  \varepsilon_2 |}{4} \frac{ \mu_{10}^4 }{\hbar^3}
e^{-(t_1 + t_3)/T_2} e^{-t_2/T_1}    \mathrm{sin}\; [ \overline{ \omega}_{01}  ( t_1 -  t_3 )  ]\; e^{ -\sigma^2 (t_1 - t_3)^2/2 },
\label{R1convolve}
\end{align}

\begin{align}
\overline{P^{(3)}_{R_4 + R_4^*} } &=  \frac{|  \varepsilon_0  \varepsilon_1  \varepsilon_2 |}{4} \frac{ \mu_{10}^4 }{\hbar^3}   e^{-(t_1 + t_3)/T_2} e^{-t_2/T_1}  \int d\omega_{01}  \Big( \frac{e ^{ - {(\omega_{01} -   
			\overline{ \omega}_{01}  )^2}  / 2 \sigma^2}}{\sigma \sqrt{2 \pi}} \Big) \;  \mathrm{sin} [ \omega_{01}( t_1 +  t_3 )  ]   \nonumber \\ &=  \frac{|  \varepsilon_0  \varepsilon_1  \varepsilon_2 |}{4} \frac{ \mu_{10}^4 }{\hbar^3}
e^{-(t_1 + t_3)/T_2} e^{-t_2/T_1}   \; \mathrm{sin} [\overline{ \omega}_{01} ( t_1 +  t_3 )  ] \; e^{ -\sigma^2 (t_1 + t_3)^2/2 }.
\label{R4convolve}
\end{align}
where  $ \overline{ \omega}_{01}  $ is the mean resonance frequency and $\sigma$ is the spread.    We plot the simulated experimental signals in Fig. \ref{TimeTracesCV2D} for both $t_1 \rightarrow 0$ and $t_2 \rightarrow 0$ (giving R, NR, and PP signals), using the same experimental values in Fig. \ref{TimeTraces2D}, but including the effects of convolution with $\sigma = 0.5$.  One can see that due to the rapid dephasing of multiple detuned oscillators the experimental signals decays quickly in the modulation directions, however one can still measure $T_1$ by the PP signal's decay in the $\hat{\tau}$ direction and $T_2$ by the R signal's decay in the $(\hat{\tau} + \hat{t})/\sqrt2$ direction.  For the non-rephasing contribution, additional decay of the experimental signal occurs as function of $t_1 + t_3$, which limits the NR signal's utility.   As can be see in Fig. \ref{TimeTracesCV2D}  (and Eq. \ref{R1convolve}) in contrast the rephasing signal shows the phenomena of ``photon echo" when $t_1 = t_3$ (e.g along the diagonal).  The important distinction between the PP and the R contributions as compared to the NR response is that they show  their decay from oscillator dephasing in the orthogonal direction from the decay that arises from lifetime effects.   In the frequency domain this will lead to a diagonal (horizontal) streaking of the signal for the R (PP) signal due to the multiple oscillators.   However the direction perpendicular to the streaking direction gives $1/T_1$ for the PP signal and $1/T_2$ for the R signal.   This is the essential important feature of 2DCS.

\subsection{Analysis}

\begin{figure*}[t]
	\includegraphics[width=11cm]{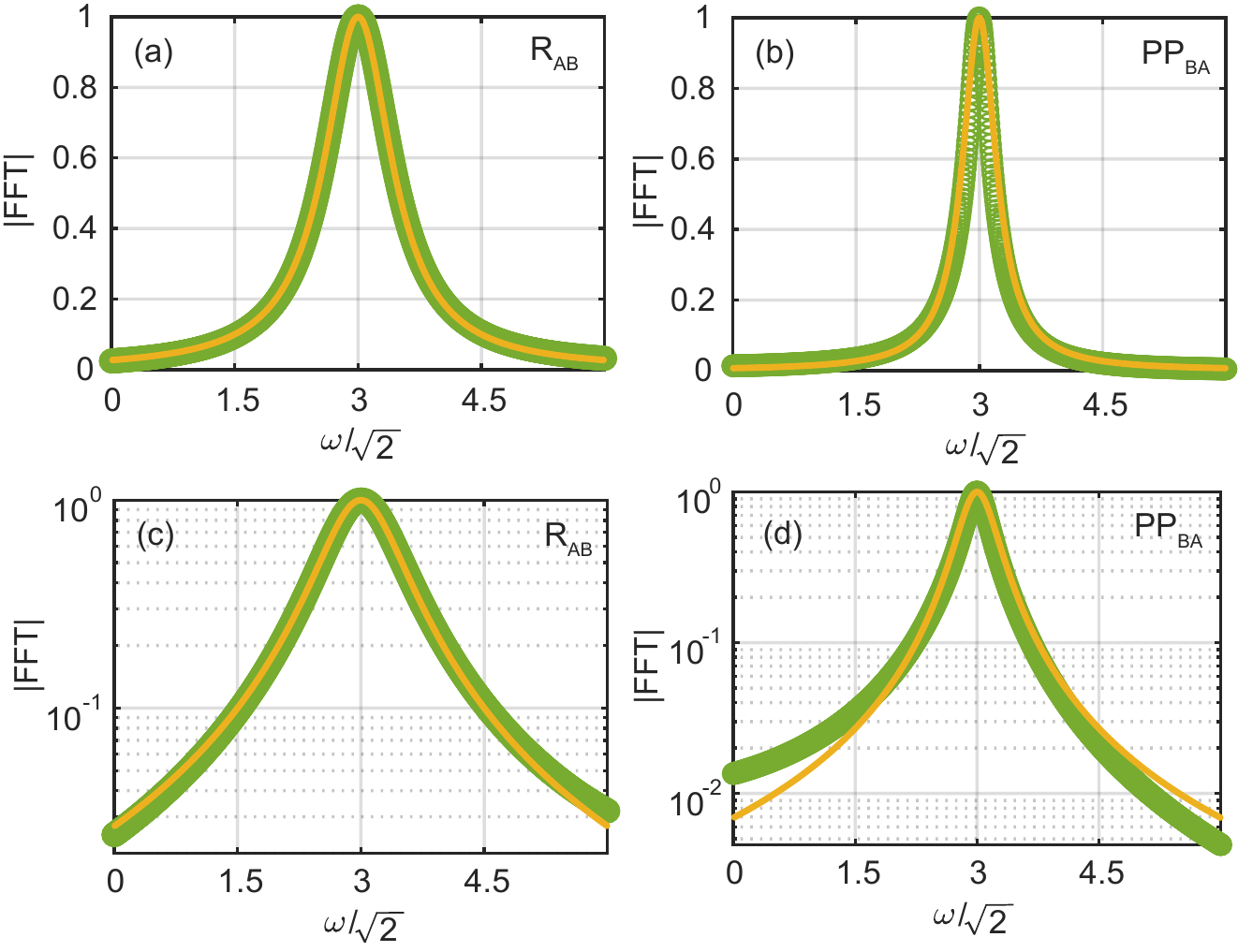}
	\caption{Magnitude of the Fourier transforms of timed-domain simulations as a function of $\omega$ showing Lorentzian lineshapes. (a) Cut along the anti-diagonal direction (yellow line in \ref{FT_2D}(c)) of the R peak for the AB pulse sequence (green dots) with the frequency axis scaled by a factor of $1/\sqrt{2}$. (b) Cut along the anti-diagonal direction (yellow line in \ref{FT_2D_BA}(a)) of the PP peak for the BA pulse sequence (green dots) also scaled by a factor of $1/\sqrt{2}$. Solid yellow lines in (a) and (b) correspond to Lorentzians with widths $1/T_2$ and $1/T_1$ respectively. (c) and (d) are the same as (a) and (b) but with a logarithmic vertical axis}
	\label{ADcuts}
\end{figure*}

Due to the ``phase twisting" in the nonlinear response functions discussed above and their dependence on both absorptive and dissipative elements it is more complicated to extract out parameter values like relaxation times than in linear response. For example, as shown in Fig. \ref{FT_2D} (a) and (b), the real and imaginary parts of the Fourier transform of the R peak along the anti-diagonal both have both positive and negative values and do not display a simple absorptive (Lorentzian) lineshape.  Also as mentioned above, the phasing procedure used in previous THz 2DCS experiments \cite{kuehn2011two} to get purely absorptive lineshapes is challenging to implement here, because of the extreme inhomogeneous broadening and overlapping contributions in 2D frequency space.  For this reason we chose to analyze the magnitudes of the experimental data in the main text as  near resonances, magnitudes are dominated by the absorptive part of the lineshape.  One can show algebraically that the magnitude of the R peak for the AB pulse sequence along the anti-diagonal direction can be approximated by a single Lorentzian with width $1/T_2$. This is illustrated in Fig. \ref{ADcuts}(a) which shows a cut along the anti-diagonal direction in the magnitude of the calculated Fourier transform of the R signal i.e., $\vert P^{(3)}_{R} \vert$ shown in Fig. \ref{FT_2D} (c). We overlay this cut with a Lorentzian  $ \propto \frac{1}{\frac{1}{T_2^2}+\left(\omega -\omega_{01}\right){}^2}$ (yellow line in Fig. \ref{ADcuts}(a)). The peak is centered at the resonance frequency $\omega_{01}$.  Note that the frequency axis (which is the anti-diagonal cut) in Fig. \ref{ADcuts} is scaled by the factor $1/\sqrt{2}$ to give it the same scaling as the $\hat{\omega}_t$ and $\hat{\omega}_\tau$ directions.  The same process is followed in analyzing the anti-diagonal cuts in the main text.  As can be seen in Fig. \ref{ADcuts}(a), the cut along the anti-diagonal is quite well described by a single Lorentzian with width $1/T_2$. Thus, to extract $1/T_2$ from the experimental data, we fit a single Lorentzian to the anti-diagonal cut across the rephasing peak over an approximately FWHM as shown in Fig. 3a of the main text. 

Similarly, it can be shown that the magnitude of the PP peak for the BA pulse sequence along it's anti-diagonal direction can be approximated by a single Lorentizan with width $1/T_1$ if one only fits the region near the peak where the response is dominated by dissipative effects. Fig. \ref{ADcuts}(b) shows a cut along the anti-diagonal direction in the magnitude of the calculated Fourier transform of the PP signal i.e., $\vert P^{(3)}_{PP} \vert$ shown in Fig. \ref{FT_2D_BA} (d). The yellow line in Fig. \ref{ADcuts}(b) is $ \propto \frac{1}{\frac{1}{T_1^2}+\left(\omega -\omega_{01}\right){}^2}$. Thus, to extract $1/T_1$ from the experimental data, we fit a single Lorentzian to the anti-diagonal cut across the PP signal for the BA pulse sequence as shown in Fig. 3a of the main text. The fits to the data in the main text were performed in a range that is approximately one FWHM so as to emphasize the dissipative character of the response.

\subsection{Broader context of 2DCS}

We have considered here only the simplest case of the 2DCS response of a two-level system.  This is a very useful example for the present case of an electron glass.  Of course most physical systems are much more complicated.  The presence of multiple oscillators with additional nonlinearities and couplings between them will result in very rich spectra.

The interested reader may ask how the above treatment can be related to the picture put forth in Ref. \cite{wan2019resolving} where the 2DCS response of an Ising chain in transverse field was calculated.   There the excited spinon pairs can be mapped to a two-level system where each pair of spinons with momenta $\pm k$ correspond to a two-level system in which the ground and excited states correspond to the absence or presence respectively of these pairs.   As discussed in the Supplemental Material to Ref. \cite{wan2019resolving} an added level of complexity occurs in that case because the spinon basis does not necessarily correspond to the same basis in which a $y$ oriented THz B field causes pure spin-flip transitions e.g. ``diagonal transitions" are also allowed.   There the Bogoliubov (energy dependent) coherence factor $\theta_k$ sets the rotation of the spinon basis from the physical reference frame in which a $y$ axis THz B field acts.   For $\theta_k = \pi/2$ (satisfied in the middle of the two spinon band), diagonal transitions are not allowed and the results of Ref. \cite{wan2019resolving} reduce to the expressions above.

We anticipate that THz 2DCS will become a powerful general tool for condensed matter physics.   For some more complicated models systems that have been analyzed previously we refer interested readers to the physical chemistry literature \cite{hamm2011concepts,Mukamel1995,hamm2005principles}.   Although there has been some attempts to work out the 2DCS response for interesting quantum magnets that exhibit fractionalization \cite{wan2019resolving,choi2020theory}, the 2D response for many materials that are at the forefront of modern condensed matter physics has not been considered at all.   This is a wide open area both theoretically and experimentally.

\section{Mechanisms for relaxation in electron glasses}

In this section, we discuss possible mechanisms for the relaxation of particle-hole excitations in the regime considered in the main text, i.e., that of an electron glass that is relatively near the metal-insulator transition.  We will proceed as follows. In Sec.~\ref{recap}, we summarize the experimental findings in the main text, and argue that they strongly constrain the plausible relaxation mechanisms.  We find that electron-electron interactions seem to be the only mechanism potentially consistent with experiment. In Sec.~\ref{dipoleham} we derive an effective model of dipolar two-level systems (TLS's), corresponding to soft particle-hole excitations. In Sec.~\ref{relaxrates} we show that dipolar interactions among the TLS's can give rise to energy relaxation with a characteristic rate $\Gamma_1 \sim \omega$. Finally we discuss (Sec.~\ref{discussion}) to what extent this mechanism can account for the experimental findings.

\subsection{Experimental constraints on possible explanations}\label{recap}

We recall the main experimental conclusions about the relaxation rates $\Gamma_1(\omega)$ and $\Gamma_2(\omega)$ for an excitation at frequency $\omega$:

\begin{enumerate}
	\item Both decay rates obey $\Gamma \sim \omega$, with a proportionality constant of order unity in the frequency range we probed ($\sim 0.2-1$ THz). 
	\item $\Gamma_1$ decreases (weakly) as the temperature is raised (i.e., relaxation slows down as the system heats up!), whereas $\Gamma_2$ is roughly temperature-independent in the range of temperatures $5 \mathrm{K} \leq T \leq 25 \mathrm{K}$. 
	\item Both decay rates decrease as the system is doped toward the metal-insulator transition.
	\item The features do not strongly on the intensity with which the system is driven.
\end{enumerate}
We also add some observations about the regime in which the experiment is performed. 

\begin{enumerate}[label=(\alph*)]
	\item A frequency of $0.5$~THz corresponds to a temperature of around $24$~K. Thus, at the lowest temperature we are probing excitations for which $\omega > k_B T$, although we move into the range $\omega \sim k_B T$ at higher $T$.
	\item The bare Coulomb interaction at the inter-phosphorus distance is approximately $3$~THz. At these temperatures (at least for the more localized samples) we are probing transitions that take place inside the Coulomb gap.
	\item The samples are doped relatively close to the metal-insulator transition \cite{Helgren02a,Helgren04a}, so the single-particle localization length is not a small parameter.
\end{enumerate}
Together, these observations imply that the experiment probes a parameter regime in which theory is not well-controlled. Nevertheless, some explanations seem implausible given the experimental results. 

\subsubsection{Rearrangements of thermally excited quasiparticles}

In short-range many-body localized systems at nonzero temperature, relaxation takes place through large-scale collective rearrangements~\cite{gopalakrishnan2015low}. The phase space for such rearrangements scales as $\omega^{-\phi}$, where $\phi \propto s(T)$, and $s(T)$ is the entropy density at temperature $T$. Since $s(T) \sim T$ at low temperatures, this mechanism would imply a temperature-dependent \emph{exponent} for the relaxation, with low-frequency relaxation being parametrically faster at higher temperatures. This mechanism is inconsistent with the data. Other mechanisms by which a TLS decays by coupling to thermal noise can be ruled out on similar grounds: noise is stronger at higher temperatures, so these mechanisms give a relaxation rate that increases with temperature, whereas we see the opposite.

\subsubsection{Phonons}

Phonons are a natural relaxation mechanism: a TLS excited can relax by emitting or absorbing a phonon. However, phonon-based mechanisms do not naturally capture the frequency- and temperature-dependence seen in the experiment. 

(i)~\emph{Acoustic phonons} give the wrong frequency- and temperature-dependence. The matrix element for a TLS at frequency $\omega$ to couple to phonons goes as $\sim \omega$~\cite{spectraldiffusion}, so the relaxation rate from coupling to phonons is $\sim \omega g(\omega) (2N({\omega},T) + 1)$, where $N(\omega, T)$ is the occupation number for bosons at frequency $\omega$ and temperature $T$. Here a factor of $N(\omega, T)$ comes from phonon absorption, and a factor of $N(\omega, T)+1$ comes from phonon emission.  Such effects cannot describe the experimental observations for two reasons. Firstly if we examine the temperature dependence we conclude that the phonon mediated relaxation rate depends on temperature according to $2N(\omega, T)+1$, which increases with temperature. However, the experimentally measured energy relaxation rate  decreases with increasing temperature. Secondly, the frequency dependence of the relaxation rate, 
in the limit when frequency is large compared to temperature, is $\omega g(\omega)$.  As the phonon density of states must vanish at $\omega \rightarrow 0$ (generally as $\omega^2$), so this will produce a relaxation rate that vanishes faster than linearly with frequency.   At low frequencies, it should go as $\omega^3$, which is inconsistent with the experimental observations~\cite{spectraldiffusion}. While this channel must exist, it seems to be subleading.

(ii)~\emph{Optical phonons} do not seem relevant as the optical phonon branch of silicon is at $15$ THz~\cite{brockhouse, beltukov}, which is well above the frequencies we are probing. In any case, the decay rate due to such phonons again would increase with temperature, due to Bose enhancement, and this is at odds with observation.

\subsubsection{Overheating}

Nonlinear current-voltage characteristics near the metal-insulator transition can exhibit bistability~\cite{altshuler2009jumps}, because phonons are ineffective at equilibrating electronic degrees of freedom to base temperature. If the electronic temperature were indeed to decouple in this way, the measured temperature-dependence would be unreliable. This scenario, however, predicts that stronger pumping should heat up the electronic degrees of freedom more. This is inconsistent with the observed insensitivity of the relaxation rates to THz intensity. 

\subsection{Effective dipolar Hamiltonian for electron glass}\label{dipoleham}

Here we start with the microscopic Hamiltonian describing impurity-band electrons in Si:P and derive a low-energy effective Hamiltonian in terms of two-level systems that interact through dipolar interactions. To allow for a controlled theoretical analysis we will assume that the localization length of single electronic excitations is short; however, we will not make any assumptions about the Coulomb interaction strength. The derivation takes place in various stages that are laid out in the following sections.

\subsubsection{From substitutional to on-site disorder}

We begin with a general tight-binding Hamiltonian for interacting electrons in the presence of positional randomness:

\begin{equation}\label{hmic}
H = \sum_{i \neq j} t_{ij} (c^\dagger_i c_j + \mathrm{h.c.}) + \frac{\mathcal{V}}{|\mathbf{r}_i - \mathbf{r}_j|} n_i n_j.
\end{equation}
Here, $\mathcal{V}$ is the characteristic coupling strength for the Coulomb interaction in this dielectric medium, $c_j$ is the annihilation operator for an electron on site $j$, and $n_j$ is the electron density on site $j$. The hopping amplitude $t_{ij} \sim \exp(-c|\mathbf{r}_i - \mathbf{r}_j|)$ falls off rapidly with the spatial separation between pairs of atoms.  The positional randomness in Eq.~\eqref{hmic} gives rise to on-site randomness through two mechanisms. First, the average potential energy on site $i$ is $\sum_j \mathcal{V}/|\mathbf{r}_i - \mathbf{r}_j| \langle n_j \rangle$, which is spatially random. If we integrate out states far from the Fermi energy, freezing in their occupation numbers, the remaining states will experience a random potential due to the randomly positioned frozen electrons. Second, even in the absence of interactions, the hopping term renormalize the on-site energy. We can evaluate the renormalized on-site energy $E_i$ self-consistently in (Brillouin-Wigner) perturbation theory. At second order we get $E_i = \sum_j \frac{|t_{ij}|^2}{E_i}$, so $
E_i = \sqrt{\sum_j |t_{ij}|^2}$. 
In practice, the $t_{ij}$ are exponentially sensitive to bond lengths so we can approximate $E_i = \max(|t_{ij}|)$, where the maximum is taken over all $j \neq i$. 
Pairs of sites that are anomalously nearby hybridize strongly (and are shifted far from the Fermi energy); perturbative corrections from coupling to these randomly positioned tightly bound dimers generate on-site disorder for typical sites.

\subsubsection{From on-site disorder to resonant pairs}

The above considerations let us replace the Hamiltonian~\eqref{hmic} with an effective Hamiltonian that describes particles on a site-diluted lattice, and contains the following terms: a random on-site potential $\epsilon_i$, drawn from a distribution of characteristic width $W$; a Coulomb interaction of strength $V_{ij} \equiv \mathcal{V}/|\mathbf{r}_i - \mathbf{r}_j|$ between any two sites; and a hopping term $t_{ij}$ of characteristic scale $t$ between neighboring pairs of sites:

\begin{equation}\label{h2}
H = \sum_{i} \epsilon_i n_i + \sum_{i\neq j} V_{ij} n_i n_j + \sum_{\langle ij \rangle} t_{ij} c^{\dag}_i c_j 
\end{equation}

We proceed as follows. If we set all the $t_{ij} = 0$ the eigenstates of Eq.~\eqref{h2} are product states in which each site is either occupied or unoccupied. The ground state of this system is the classical ``electron glass''.  At low temperatures, there are many metastable states, all with statistically similar properties that the system is in some mixture of. The ground state and metastable states are, by definition, stable against single-electron moves.  This stability implies a pseudogap (Coulomb gap) in the density of states~\cite{Efros75a}. 

Starting from this classical ground state, we would like to create a particle-hole excitation at a low frequency $\omega \ll W, t$, by moving an electron from a filled level $a$ to an empty level $b$. This dimer has four states, which we label $|00\rangle$ (neither filled), $|\downarrow\rangle$ (only $a$ filled, i.e., ground state), $|\uparrow\rangle$ (only $b$ filled) and $|11\rangle$ (both filled). We ignore the $|00\rangle$ state in what follows. The remaining three states have the following energies:

\begin{equation}\label{SEcount}
E_{\downarrow} = \epsilon_a + \sum_{c \neq a, b} V_{ac}, \quad E_{\uparrow} = \epsilon_b + \sum_{c \neq a,b} V_{bc}, \quad E_{11} = E_{\downarrow} + E_{\uparrow} + V_{ab}.
\end{equation}

A pump at frequency $\omega$ can induce transitions between pairs of levels with energies separated by $\omega$ (up to some resolution set by the properties of the pulse). The matrix element connecting states $|\downarrow\rangle$ and $ |\uparrow\rangle$ comes from their hybridization due to the hopping. In what follows it is crucial to understand how the hopping shifts these levels. The matrix element goes as $t_{ab} = W e^{-|r_{ab}|/\xi}$ where $\xi$ is the localization length. (Within the locator approximation, $t_{ab}$ is due to a sequence of $\sim |r_{ab}|$ non-degenerate perturbative hops, so $t_{ab} \sim W(t/W)^{|r_{ab}|}$, implying that $\xi \sim 1/\log(W/t)$. Once this hybridization is included, the splitting between the two levels is

\begin{equation}
\mathcal{E}_{ab} = \sqrt{t_{ab}^2 + (E_{\uparrow} - E_{\downarrow})^2}.
\end{equation}

The light can excite a transition when its frequency $\omega$ matches the energy of the energy splitting.  This condition can not hold unless $t_{ab} \alt \omega$. Therefore, there is a minimal distance $r_\omega \sim \xi \log(W/\omega)$ above which such resonances are possible~\cite{Mott79a}. Moreover, if $r \gg r_\omega$, hopping is very unlikely (i.e., has probability $\sim e^{-r/r_\omega}$) to appreciably hybridize the levels, and the transition amplitude is strongly suppressed. Therefore, as first pointed out by Mott, resonant pairs at splitting $\omega$ form at an optimal distance $r_\omega$~\cite{Mott79a}. At this optimal scale, the states $|\uparrow\rangle, |\downarrow\rangle$ of the resonant pair are strongly hybridized, and form a TLS governed by the Hamiltonian

\begin{equation}\label{htls}
H_{\mathrm{TLS}} = t_{ab} \sigma^x_{ab} + (E_{\downarrow} - E_{\uparrow}) \sigma^z_{ab} \equiv \mathcal{E}_{ab} \tau^z_{ab}.
\end{equation}
In what follows, we call such pairs of levels resonant pairs. We will next estimate the density of such resonant pairs. 

\subsubsection{Density of active resonant pairs}

To contribute to low-frequency dynamics, a resonant pair must be ``active'' in the ground state, i.e., it should be in the states $|\downarrow\rangle$ or $|\uparrow\rangle$. We now estimate the density of pairs that meet this criterion. There are two regimes:

\begin{description}
	\item[Mott regime] When $\omega$ is relatively large, i.e., $\omega \agt \mathcal{V}/|r_\omega|$, the interaction correction to $E_{11}$ in Eq.~\eqref{SEcount} is unimportant. Thus, for a resonant pair to be active, the lower eigenstate of the TLS must be within $\omega$ of the Fermi energy. The density of resonant pairs at frequency $\omega$ then scales as $\omega r_\omega^2$.
	\item[Shklovskii-Efros regime] When $\omega \alt \mathcal{V}/|r_\omega|$, the interaction correction in Eq.~\eqref{SEcount} crucially changes the counting, through a mechanism similar to Coulomb blockade. As long as \emph{either} $E_0$ or $E_1$ is within $\mathcal{V}/|r_\omega|$ of the Fermi energy, the $|11\rangle$ state is energetically unfavorable, so the pair of sites is singly occupied in the ground state. Therefore, the phase space for the TLS goes as $\mathcal{V}|r_\omega|$, which is essentially constant at low frequencies~\cite{shklovskii1981phononless}. 
\end{description}

The crossover between these two regimes is also captured by the low-frequency (phononless) conductivity, which goes as $\omega^2$ in the Mott regime and $\omega$ in the Shklovskii-Efros regime~\cite{shklovskii1981phononless}.

\subsubsection{Interactions between resonant pairs}

We now argue that the Coulomb interaction induces effective dipolar ($1/R^3$) interactions between resonant pairs. The result is well known~\cite{burin1994low,burin,yao2014many}, but we rederive it for completeness.  We consider two active resonant pairs $ab$ and $cd$, separated by a distance that is much larger than the size of the dipoles themselves i.e. $r_{ac} \gg \text{max}(r_{ab}, r_{cd})$.   Since both pairs are assumed to be active, the two-pair Hamiltonian acts in a four-dimensional space consisting of the two TLS's $\mathbf{\sigma}_{ab}$ and $\mathbf{\sigma}_{cd}$. The Coulomb interaction is diagonal in the $\sigma^z$ basis for both TLSs. For example, if the pair $ab$ is in its $|\downarrow\rangle$ state, the potential on site $c$ is given by $V_0 + V_{ac}$, where $V_0$ is the potential due to all spins other than the resonant pairs under consideration; similarly, the potential on site $d$ in this configuration is $V_0 + V_{ad}$. Writing this out, we see that the Coulomb interaction between the two resonant pairs takes the form

\begin{equation}
H_C = \frac{1}{4}(1 - \sigma^z_{ab}) (V_{ac} - V_{ad}) \sigma^z_{cd} + \frac{1}{4}(1 + \sigma^z_{ab}) (V_{bc} - V_{bd}) \sigma^z_{cd} = \frac{1}{2} (V_{bc} - V_{ac} + V_{ad} - V_{bd}) \sigma^z_{ab} \sigma^z_{cd} + \ldots
\end{equation}
where the terms ignored in $\ldots$ are overall static shifts that are included in the full Hamiltonian. Assuming $r_{ac} \gg \mathrm{max}(r_{ab}, r_{cd})$, one can expand the effective coupling constant in a multipole expansion to get that $V_{ab, cd} \sim \mathcal{V} r_{ab} r_{cd} / r_{ac}^3$. (In this expression we have ignored some geometric factors that will not be important in what follows.) Thus, the Hamiltonian for two coupled resonant pairs is

\begin{equation}\label{4lev}
H_{\mathrm{2-pair}} = \Delta_{ab} \sigma^z_{ab} + t_{ab} \sigma^x_{ab} + \Delta_{cd} \sigma^z_{cd} + t_{cd} \sigma^x_{cd} + \frac{c'\mathcal{V} r_{ab} r_{cd}}{r_{ac}^3} \sigma^z_{ab} \sigma^z_{cd}.
\end{equation}
where $c'$ is a numerical prefactor. 
In the regime of interest, we can simplify this further. The Hamiltonian in Eq.~\eqref{4lev} describes a four-dimensional Hilbert space. However, when the Coulomb interaction between the two resonant pairs is weaker than (or comparable to) their splitting, we can make a secular approximation and eliminate the matrix elements of the Coulomb interaction that do not preserve the number of excitations. Doing so, and transforming to the eigenbasis of the isolated TLSs (denoted by Pauli matrices $\tau$), we get the equation for two interacting resonant pairs:

\begin{equation}\label{secular}
H_{\mathrm{2-pair}} = \mathcal{E}_{ab} \tau^z_{ab} + \mathcal{E}_{cd} \tau^z_{cd} + \frac{\tilde c \mathcal{V} r_{ab} r_{cd}}{r_{ac}^3} (\tau^+_{ab} \tau^-_{cd} + \mathrm{h.c.})
\end{equation}
where $\tilde c\neq c'$ is a prefactor that is not important for our argument. In what follows we will consider relaxation in an ensemble of resonant pairs interacting pairwise through Eq.~\eqref{secular}.

\subsection{Relaxation of coupled resonant pairs}\label{relaxrates}

After eliminating the degrees of freedom that are not low-frequency resonant pairs, we finally reduce the Hamiltonian~\eqref{hmic} to a spin model for coupled TLS's, of the form

\begin{equation}\label{cpair}
H_{\mathrm{pair}} = \sum_\alpha \mathcal{E}_\alpha \tau^z_\alpha + \sum_{\alpha\neq \beta} \frac{ c|\mathcal{V}| p_\alpha p_\beta}{r_{\alpha\beta}^3} (\tau^+_\alpha \tau^-_\beta + \mathrm{h.c.}). 
\end{equation}
where $ c$ is an prefactor that is not important for the argument,  $\alpha$ labels resonant pairs, and $p_\alpha \sim \xi\log(W/\mathcal{E}_\alpha)$ is the size of pair $\alpha$. In the dilute limit and when all pairs but one are in their ground state, Eq.~\eqref{cpair} describes a single particle hopping on a disordered lattice with dipolar interactions, a problem addressed in Ref.~\cite{levitov1989absence}. The density of resonant pairs depends on which regime one considers:

\begin{equation}
P(\mathcal{E}_\alpha = \omega) \sim \frac{1}{W^2} \left\{
\begin{array}{lr}
\mathcal{V} \xi \log(W/\omega) & \text{Shklovskii-Efros} \\
& \\
\omega \xi^2 \log^2(W/\omega) & \text{Mott}
\end{array}\right.
\end{equation}
We will deal with these cases separately.

\subsubsection{Shklovskii-Efros regime}

The experiment flips a pseudospin at frequency $\omega$ and we are interested in how the pseudospin decays. We can eliminate all states in the Hilbert space of Eq.~\eqref{cpair} with energies greater than $c \omega$, where $c$ is some constant of order unity. (One can regard this as decoupling such states through a Schrieffer-Wolff transformation; the transformation will generate additional short-range interactions that do not affect the analysis.) This leaves behind a sparse network of effective spins, with characteristic bandwidth $c \omega$ and spatial separation   $[c \omega \mathcal{V} \xi \log(W/\omega) / W^2]^{-1/3}$. The interaction between typical spins at this scale goes as $V_{\mathrm{eff}} \sim \omega \xi^3 \log^3(W/\omega) (\mathcal{V}/W)^2$. Thus, a dipole excited by a frequency $\omega$ typically has a nearest neighbor at a distance where the interaction is resonant.   

Note that this occurs precisely because the dipolar interaction scales as inverse volume in three dimensions i.e. it is marginal~\cite{levitov1989absence}. More generally, if interactions fall off as $1/R^\alpha$ we would get $V_{\mathrm{eff}} \sim \omega^{d/\alpha}$, so the disorder would become irrelevant (for $\alpha < d$) or the interactions would become irrelevant (for $d < \alpha$) giving rise to dipole localization. 

Since the rate at which TLS's hop on the effective lattice scales as $V_{\mathrm{eff}}$, this mechanism would predict 

\begin{equation}\label{lse}
\Gamma(\omega) \simeq \omega \xi^3 \log^3(W/\omega) (\mathcal{V}/W)^2.    
\end{equation}
This $\omega$-dependence is consistent with experimental observations, though it is not obvious why the approximations we have made to derive Eq.~\eqref{lse} should be valid for the parameters in the experiment. In particular, our calculation has assumed that the localization length $\xi$ is small compared to the $P-P$ spacing, that the `dipoles' are well separated (inter-dipole spacing much larger than dipole size), and that the frequency $\omega$ is much smaller than the characteristic disorder and interaction scales. These assumptions are suitable deep in the insulating phase, but none of these assumptions are safe in the experimental regime close to the MIT, where $\xi$ is large compared to $P-P$ spacing, and where $\omega$ is of the same order as the characteristic disorder and interaction scales (such that the inter-dipole separation is comparable to the dipole size). These issues are explored further in the discussion section below. 

\subsubsection{Mott regime}

Adapting the previous discussion to the Mott regime is straightforward. The spatial density of low-energy dipoles now scales as $\omega^2$ rather than $\omega$. Thus the spacing of low-energy pairs therefore scales as $\omega^{-2/3}$, and the interactions at that scale go as $\omega^2$. In this case, therefore, the interactions are parametrically weaker than the hopping of dipoles on the effective lattice. Naively one might estimate that $\Gamma(\omega) \sim \omega^2$.  Resonant pairs below some frequency scale have no neighbors to inter-resonate with; they delocalize instead through the long-range part of the dipole interaction~\cite{levitov1989absence}, with relaxation rate that is therefore much slower.  So unlike the previous case these dipoles do get arbitrarily sharp as $\omega \rightarrow 0$, which implies such a system would be a “Fermi glass”.

\subsubsection{Short-range interactions}

Although not likely relevant to experiment, it is illustrative to apply the reasoning above to the case of short-range interactions. Once again, the density of active resonant pairs at frequency $\omega$ scales as a power of $\omega$ (which depends on whether $\omega$ is in the Shklovskii-Efros or Mott regimes). The typical spacing between dipoles is thus $\omega^{-\phi}$ for $\phi = 1/3$ (Shklovskii-Efros regime) or $2/3$ (Mott regime). However, the matrix element between dipoles at this scale is now exponentially suppressed, and scales as $\sim \exp(-1/\omega^\phi) \ll \omega$. A naive application of the Golden Rule might suggest that the lifetime should scale similarly, yielding arbitrarily sharp excitations near the Fermi energy. However, in this case we expect that a typical dipole has \emph{no} resonant neighbors, and should therefore be strictly localized, below some critical frequency.

\subsubsection{An alternative derivation}

The discussion above involved some heuristic steps, so it is helpful to check it using a more systematic approach. To this end we adopt the self-consistent theory of localization~\cite{abou1973selfconsistent}, as follows. We compute the lifetime $\Gamma(\omega)$ of a dipole due to hopping to other sites on the lattice, while assuming that these other sites also decay at rate $\Gamma(\omega)$. A self-consistent approach is necessary because otherwise each line in a localized system is infinitely sharp, and exact resonances cannot be found. To keep the model tractable we assume that the decay rate is only a function of $\omega$ (i.e., we neglect spatial heterogeneity that could in principle be important). This leads us to a self-consistent equation for the function $\Gamma(\omega)$:

\begin{equation}
\Gamma_\alpha = \sum_\beta V_{\alpha\beta}^2 \mathrm{Im} \left( \frac{1}{\omega_\alpha - \omega_\beta + i \Gamma(\omega_\beta)} \right).
\end{equation}
Converting the sum to an integral and invoking spatial homogeneity we get

\begin{equation}
\Gamma(\omega) = \int_{R(\omega)}^\infty dr r^2 \frac{\mathcal{V}^2}{r^6} \int d\Omega \frac{\mathcal{V} \xi \log(W/\Omega)}{W^2} \frac{\Gamma(\Omega)}{(\omega - \Omega)^2 + \Gamma(\Omega)^2}.
\end{equation}
Here we have used the crucial fact that resonant pairs at $\omega$ cannot be spaced much closer than $R(\omega) \sim (\mathcal{V}/\omega)^{1/3}$, as otherwise the dipolar interaction would hybridize and split them strongly. Outside this zone, the detunings and frequencies are uncorrelated, allowing us to simplify this expression (neglecting logarithmic factors) to be

\begin{equation}
\Gamma(\omega) = \omega \frac{\mathcal{V}^2}{W^2}  \int d\Omega \frac{\Gamma(\Omega)}{(\omega - \Omega)^2 + \Gamma(\Omega)^2}.
\end{equation}
One can check that the only power-law scaling that satisfies this self-consistent equation is $\Gamma \sim \omega (\mathcal{V}/W)^2$, confirming our previous estimate up to logarithmic factors that have been suppressed.

\subsection{Discussion}\label{discussion}

The analysis of the previous section shows how a combination of Coulomb blockade and dipolar hopping can give rise to the experimentally observed $\Gamma(\omega) \sim \omega$ for the relaxation of a TLS in an electron glass. The prefactor (ignoring logarithms) is $\xi^3 (\mathcal{V}/W)^2$, which is not necessarily small, if the typical microscopic hopping amplitude between adjacent P sites is comparable to the Coulomb interaction between them. This framework also naturally reproduces the sharpening of the TLS frequency as one increases the temperature up to $T \simeq \omega$: when delocalization takes place through coherent tunneling at zero temperature, the finite-temperature corrections usually suppress transport through decoherence (see, e.g., Ref.~\cite{fisherzwerger}), and furthermore raising temperature increases screening and hence suppresses the Coulomb interaction. 

As mentioned above, the validity of our analysis relied on the assumptions that $\xi$ is small compared to the $P-P$ spacing and $\omega$ is small compared to the typical hopping and interaction scale. These assumptions are not quantitatively accurate in the experimental regime (close to the metal insulator transition), where $\xi$ becomes large compared to the $P-P$ spacing, and $\omega$ is of the same order as the typical interaction scale between neighboring $P$ dopants.   We can briefly comment on how some version of this mechanism might operate closer to the metal-insulator transition.  There are two essential ingredients: (i) hopping of dipolar excitations among electron-hole pairs and (ii) the enhancement of phase space for active resonant pairs due to Coulomb blockade. As one approaches the metal-insulator transition, the picture of well-separated dipoles breaks down: rather, one has a much more densely connected network of overlapping electron-hole pairs. Nevertheless, an excitation at frequency $\omega$ can only hybridize with states that are at frequency $\simeq \omega$, and (by Mott's argument) the matrix element for this cannot exceed $\omega$. Thus, the $\omega$-dependence of decay rates should persist. 
As $\xi$ increases past unity, however, the Coulomb blockade effect becomes less effective; naively, one should replace $\mathcal{V}$ in Eq.~\eqref{lse} by the charging energy $\mathcal{V}/\xi^3$. Also, the bandwidth $W$ is increasingly dominated by kinetic energy, so $\mathcal{V}/W$ decreases as a consequence. These considerations are consistent with the dimensional argument in the main text. 
Thus, dipolar relaxation seems consistent with the observed trends near the metal-insulator transition, though we have not been able to perform a controlled calculation in that regime.

Our discussion so far has been limited to estimating the lifetime of a single particle-hole excitation in an electron glass (i.e., the $T_1$ time). We now turn to two other issues: the relation between the dipole lifetime and the lifetime of the single-particle Green's function, and the relation between $T_1$ and $T_2$ times. 

\subsubsection{Electron lifetime vs. dipole lifetime}

The calculation outlined above gives the lifetime of an elementary dipole in an electron glass. The question of whether a quasiparticle is sharp, however, is normally phrased in terms of the decay rate of the single-electron Green's function, probed, for example, by tunneling an electron into the system. In translationally invariant Fermi liquids the two quantities are effectively the same; momentum and energy conservation forbid the particle from recombining with its hole, so they relax separately as if they were injected excitations. In the Fermi glass, by contrast, the lack of momentum conservation allows the particle and hole to recombine. 

We now briefly comment on what happens if one injects an electron into the electron glass. In the Mott regime, this calculation is straightforward; the electron relaxes into two particles and a hole. Since all three final states must lie within $\omega$ of the Fermi level, and there are two free energies, the phase space scales as $\omega^2$. Therefore the lifetime of an injected particle is parametrically longer than that of a dipole: it scales as $\omega^{-3}$. %
In the Shklovskii-Efros regime this question is much more delicate, owing to the presence of the Coulomb gap: we do not address this question here, but remark that a similar separation of scales seems possible. As noted in the main text, we are treating the particle-hole lifetime rather than the injected-particle lifetime as the fundamental quantity, since this is the relevant low energy excitation for an isolated system. 

\subsubsection{Thermal effects, spectral diffusion, and $T_1$ vs. $T_2$}

Within our zero-temperature theory the processes responsible for $T_1$ and $T_2$ are the same, so we expect these quantities to be related by some simple scale factor; this is indeed what we observe. However, the trends with increasing temperature are different: $T_1$ grows with temperature, while $T_2$ remains roughly constant (so the ratio $T_1/T_2$ increases). A natural interpretation is that this approximate temperature-independence comes from a competition between the increasing $T_1$ time and the opening of thermal dephasing channels. 

Another potential relaxation channel is spectral diffusion~\cite{spectraldiffusion}. The essential idea behind spectral diffusion is that the splitting of a particular TLS is a time-varying quantity, undergoing stochastic fluctuations with ``amplitude" $\Delta \omega$ and correlation time $\tau$.  When the correlation time is sufficiently short, we expect that the energy splitting of a TLS fluctuates at a rate $\sim \Delta \omega^2 \tau$ \cite{hamm2011concepts}. Note that this rate decreases as $\tau$ gets shorter: this is analogous to ``motional narrowing,'' where rapid fluctuations in the environment get averaged out. If $\tau$ were to decrease with temperature, while $\Delta \omega$ stayed roughly constant, spectral diffusion may give rise to lifetimes that increased with temperature. As we will discuss below, the natural physical mechanisms for spectral diffusion have the frequency amplitude increasing as its correlation time decreases, so the predicted lifetime shortens as the system is heated up.

Two possible mechanisms for noise on a particular resonant pair are phonons and dipolar fluctuations of other resonant pairs. 
As we have already discussed, effects due to phonons should be strongly suppressed at low frequency and temperature, in a manner inconsistent with our observations.  We now turn to interactions among resonant pairs.
A given resonant pair is surrounded by thermally fluctuating pairs with typical energy (and lifetime) $\sim T$.  When $\hbar \omega \alt k_B T$, each such fluctuation is large compared with the splitting of the TLS; therefore, in this regime one cannot think of a TLS as having a stable splitting $\mathcal{E}$. Rather one must work in the $z$ basis of Eq.~\eqref{htls}, and consider incoherent transitions due to the hopping. In this regime, any specific level fluctuates over a range $\sim T$ on a timescale $\sim 1/T$, so the decay rate is not strongly frequency dependent but scales as $T$; this is inconsistent with our observations. Another way to see this is as follows: the lifetime of a dipole at temperature $T$ scales as $1/T$, but because the interaction scales as inverse volume, the strength of the noise generated by nearby dipoles at temperature $T$ also scales as $T$, so $\Delta \omega^2 \tau \sim T$.

In addition to energy decay, spectral diffusion potentially gives rise to dephasing. This is because the rate of phase accumulation depends on the energy of the excitation, and if the energy is fluctuating about some mean, then the rate of phase accumulation is also fluctuating. Very roughly, fluctuations in the energy of the excitation lead to a random walk in the phase, and by the central limit theorem, if the line broadening from spectral diffusion is $\sim k_B T$, then the rate of dephasing from spectral diffusion goes $\sim \sqrt{k_B T}$.

It is worth emphasizing that the line broadening and the dephasing rate arising from spectral diffusion are both {\it increasing} functions of temperature. In contrast, we measure that the energy relaxation rate is a decreasing function of temperature, and the dephasing rate is approximately temperature independent.   Although it could be that more complicated models of spectral diffusion that incorporate the effects of temperature dependent screening do describe the data, simple models of spectral diffusion do not describe our data.   We will also mention that despite these arguments, it could still be quite interesting to try to measure the spectral diffusion directly via the multi-pulse 2DCS experiments \cite{hamm2005principles}.   Accepted protocols for measuring spectral diffusion exist in molecular systems that could be adapted to the THz 2DCS.   These will be topics for further study.     

\subsection{Summary}

We end this discussion of relaxation rates with a brief summary. Many natural potential mechanisms---such as thermal noise, spectral diffusion, Joule heating, and electron-phonon interactions---do not seem to be relevant to the physics in the regimes we are probing. In some cases this can be seen a priori, in other cases because these mechanisms do not fit the data. The remaining candidate is the Coulomb interaction. We know that Coulomb interactions affect the response of the system in the frequency range being probed (this is evident, for example, in the Shklovskii-Efros dependence of conductivity in Fig.~2 of the main text). We argued that these interactions give rise to relaxation via a continuum of dipoles, made up of localized pairs of single-particle orbitals; the relaxation rate goes as $\Gamma_1 \sim \omega$, as observed in the experiment. This dipolar theory (developed for low frequency response in deeply localized systems) is not well controlled in the experimental parameter regime, which probes intermediate frequencies in systems near the metal-insulator transition.  However, it is consistent with the experimental phenomenology at low temperatures. Extending this theory of dipolar relaxation near the metal-insulator transition and to finite temperatures is an interesting task for future work.

\end{document}